\documentclass[useAMS,usenatbib]{mnras}

\usepackage{graphicx}
\usepackage{graphics}
\usepackage{color}
\usepackage{url}
\usepackage[caption=false]{subfig}
\usepackage{verbatim}
\usepackage{array}
\usepackage{amssymb}
\usepackage{amsmath}
\usepackage{afterpage}
\usepackage{alphalph}

\usepackage{lscape}
\usepackage{pbox}
\usepackage{supertabular}

\newcommand{\swift}{{\it Swift}}

\begin{document}

\title[ASAS-SN Bright SN Catalog 2015]{The ASAS-SN Bright Supernova Catalog -- II. 2015}

\author[T.~W.-S.~Holoien et al.]{T.~W.-S.~Holoien$^{1,2,3}$\thanks{tholoien@astronomy.ohio-state.edu}, J.~S.~Brown$^{1}$, K.~Z.~Stanek$^{1,2}$, C.~S.~Kochanek$^{1,2}$,  
\newauthor
B.~J.~Shappee$^{4,5}$, J.~L.~Prieto$^{6,7}$, Subo~Dong$^{8}$, J.~Brimacombe$^{9}$, D.~W.~Bishop$^{10}$,     
\newauthor
U.~Basu$^{1,11}$, J.~F.~Beacom$^{1,2,12}$, D.~Bersier$^{13}$, Ping~Chen$^{8}$, A.~B.~Danilet$^{12}$,   
\newauthor 
E.~Falco$^{14}$, D.~Godoy-Rivera$^{1}$, N.~Goss$^{1}$, G.~Pojmanski$^{15}$, G.~V.~Simonian$^{1}$,  
\newauthor
D.~M.~Skowron$^{15}$, Todd~A.~Thompson$^{1,2}$, P.~R.~Wo\'zniak$^{16}$, C.~G.~\'{A}vila$^{17}$,  
\newauthor
G.~Bock$^{18}$, J.-L.~G.~Carballo$^{19}$, E.~Conseil$^{20}$, C.~Contreras$^{17}$, I.~Cruz$^{21}$,  
\newauthor
J.~M.~F.~And\'ujar$^{22}$, Zhen~Guo$^{8,23}$, E.~Y.~Hsiao$^{24}$, S.~Kiyota$^{25}$, R.~A.~Koff$^{26}$, 
\newauthor
G. Krannich$^{27}$, B.~F.~Madore$^{4}$, P.~Marples$^{28}$, G.~Masi$^{29}$, N.~Morrell$^{17}$, 
\newauthor
L.~A.~G.~Monard$^{30}$, J.~C.~Munoz-Mateos$^{31}$, B.~Nicholls$^{32}$, J.~Nicolas$^{33}$,  
\newauthor
R.~M.~Wagner$^{1,34}$, and W.~S.~Wiethoff$^{35}$\\ \\
  $^{1}$ Department of Astronomy, The Ohio State University, 140 West 18th Avenue, Columbus, OH 43210, USA \\
  $^{2}$ Center for Cosmology and AstroParticle Physics (CCAPP), The Ohio State University, 191 W. Woodruff Ave., \\
             \hspace{0.6cm}Columbus, OH 43210, USA \\
  $^{3}$ US Department of Energy Computational Science Graduate Fellow \\
  $^{4}$ Carnegie Observatories, 813 Santa Barbara Street, Pasadena, CA 91101, USA \\
  $^{5}$ Hubble and Carnegie-Princeton Fellow\\
  $^{6}$ N\'ucleo de Astronom\'ia de la Facultad de Ingenier\'ia, Universidad Diego Portales, Av. Ej\'ercito 441, Santiago, Chile \\
  $^{7}$ Millennium Institute of Astrophysics, Santiago, Chile \\
  $^{8}$ Kavli Institute for Astronomy and Astrophysics, Peking University, Yi He Yuan Road 5, Hai Dian District, \\
               \hspace{0.6cm}Beijing 100871, China \\
  $^{9}$ Coral Towers Observatory, Cairns, Queensland 4870, Australia \\
  $^{10}$ Rochester Academy of Science, 1194 West Avenue, Hilton, NY, 14468, USA \\
  $^{11}$ Grove City High School, 4665 Hoover Road, Grove City, OH 43123, USA \\
  $^{12}$ Department of Physics, The Ohio State University, 191 W. Woodruff Ave., Columbus, OH 43210, USA \\
  $^{13}$ Astrophysics Research Institute, Liverpool John Moores University, 146 Brownlow Hill, Liverpool L3 5RF, UK \\
  $^{14}$ Harvard-Smithsonian Center for Astrophysics, 60 Garden St., Cambridge, MA 02138, USA \\
  $^{15}$ Warsaw University Astronomical Observatory, Al. Ujazdowskie 4, 00-478 Warsaw, Poland \\
  $^{16}$ Los Alamos National Laboratory, Mail Stop B244, Los Alamos, NM 87545, USA \\
  $^{17}$ Las Campanas Observatory, Carnegie Observatories, Casilla 601, La Serena, Chile \\
  $^{18}$ Runaway Bay Observatory, 1 Lee Road, Runaway Bay, Queensland 4216, Australia \\
  $^{19}$ Observatorio Cerro del Viento, MPC I84, Pl. Fernandez Pirfano 3-5A, 06010 Badajoz, Spain \\
  $^{20}$ Association Francaise des Observateurs d'Etoiles Variables (AFOEV), Observatoire de Strasbourg, 11 Rue de l'Universite, \\
               \hspace{0.6cm}67000 Strasbourg, France \\
  $^{21}$ Cruz Observatory, 1971 Haverton Drive, Reynoldsburg, OH, 43068, USA \\
  $^{22}$ Observatory Inmaculada del Molino, Hernando de Esturmio 46, Osuna, 41640 Sevilla, Spain \\
  $^{23}$ Department of Astronomy, Peking University, Yi He Yuan Road 5, Hai Dian District, Beijing 100871, China \\
  $^{24}$ Department of Physics, Florida State University, 77 Chieftain Way, Tallahassee, FL, 32306, USA \\
  $^{25}$ Variable Star Observers League in Japan, 7-1 Kitahatsutomi, Kamagaya, Chiba 273-0126, Japan \\
  $^{26}$ Antelope Hills Observatory, 980 Antelope Drive West, Bennett, CO, 80102, USA \\
  $^{27}$ Roof Observatory Kaufering, Lessingstr. 16, D-86916 Kaufering, Germany \\
  $^{28}$ Leyburn \& Loganholme Observatories, 45 Kiewa Drive, Loganholme, Queensland 4129, Australia \\
  $^{29}$ Virtual Telescope Project, Via Madonna de Loco, 47-03023 Ceccano (FR), Italy \\
  $^{30}$ Kleinkaroo Observatory, Calitzdorp, St. Helena 1B, P.O. Box 281, 6660 Calitzdorp, Western Cape, South Africa \\
  $^{31}$ European Southern Observatory, Casilla 19001, Santiago, Chile \\
  $^{32}$ Mount Vernon Observatory, 6 Mount Vernon Place, Nelson, New Zealand \\
  $^{33}$ Groupe SNAude France, 364 Chemin de Notre Dame, 06220 Vallauris, France \\
  $^{34}$ LBT Observatory, University of Arizona, Tucson, AZ 85721-0065 USA \\
  $^{35}$ Department of Earth and Environmental Sciences, University of Minnesota, 230 Heller Hall, 1114 Kirby Drive, \\
               \hspace{0.6cm}Duluth, MN. 55812, USA
  }
\maketitle
\begin{abstract}
This manuscript presents information for all supernovae discovered by the All-Sky Automated Survey for SuperNovae (ASAS-SN) during 2015, its second full year of operations. The same information is presented for bright ($m_V\leq17$), spectroscopically confirmed supernovae discovered by other sources in 2015. As with the first ASAS-SN bright supernova catalog, we also present redshifts and near-UV through IR magnitudes for all supernova host galaxies in both samples. Combined with our previous catalog, this work comprises a complete catalog of 455 supernovae from multiple professional and amateur sources, allowing for population studies that were previously impossible. This is the second of a series of yearly papers on bright supernovae and their hosts from the ASAS-SN team.
\end{abstract}

\raggedbottom


\section{Introduction}
\label{sec:intro}

In recent decades, systematic searches for supernovae have progressed into the era of large projects that survey some or all of the sky for supernovae and other transient phenomena using varying degrees of automation. These projects include the Lick Observatory Supernova Search \citep[LOSS;][]{li00}, the Panoramic Survey Telescope \& Rapid Response System \citep[Pan-STARRRS;][]{chambers16}, the Texas Supernova Search \citep{quimby06}, the Sloan Digital Sky Survey (SDSS) Supernova Survey \citep[][]{frieman08}, the Catalina Real-Time Transient Survey \citep[CRTS;][]{drake09}, the CHilean Automatic Supernova sEarch \citep[CHASE;][]{pignata09}, the Palomar Transient Factory \citep[PTF;][]{law09}, the Gaia transient survey \citep{hodgkin13}, the La Silla-QUEST (LSQ) Low Redshift Supernova Survey \citep{baltay13}, the Mobile Astronomical System of TElescope Robots \citep[MASTER;][]{gorbovskoy13} survey, the Optical Gravitational Lensing Experiment-IV \citep[OGLE-IV;][]{wyrzykowski14}, and, recently, the Asteroid Terrestrial-impact Last Alert System \citep[ATLAS;][]{tonry11}, among numerous others. 

Prior to 2013, however, there was no rapid-cadence optical survey that surveyed the entire visible night sky to find the bright, nearby supernovae that can be studied most comprehensively, and thus have the largest impact on our understanding of supernova physics. This changed with the creation of the All-Sky Automated Survey for SuperNovae (ASAS-SN\footnote{\url{http://www.astronomy.ohio-state.edu/~assassin/}}; \citealt{shappee14}), a long-term project designed to find bright transients, such as nearby supernovae \citep[e.g.,][]{dong16,holoien16c,shappee16}, tidal disruption events \citep{holoien14b,brown16a,brown16b,holoien16b,holoien16a}, flares in active galactic nuclei \citep{shappee14}, stellar outbursts \citep{holoien14a,schmidt14,herczeg16,schmidt16}, and cataclysmic variable stars \citep{kato13,kato14,kato15,kato16}. 

ASAS-SN accomplishes this using 8 telescopes with 14-cm aperture lenses and standard $V$-band filters which have a $4.5\times4.5$ degree field-of-view and a limiting magnitude of $m_V\sim17$ (see \citet{shappee14} for further technical details). The ASAS-SN telescopes are divided into two units, each consisting of four telescopes on a common mount hosted by the Las Cumbres Observatory Global Telescope Network \citep[LCOGT;][]{brown13}. Our northern unit, Brutus, is hosted at the LCOGT site on Mount Haleakala in Hawaii, while our southern unit, Cassius, is hosted at the LCOGT site at Cerro Tololo, Chile. These two units give us roughly 20000 square degrees of coverage per clear night, and we cover the entire observable sky ($\sim30000$ square degrees on a given night) with a $2-3$ day cadence. All data are processed and automatically searched in real-time, which allows for rapid discovery and response, and discoveries are announced publicly, sometimes within a few hours of data collection. For a more detailed history of the ASAS-SN project, see the introduction of \citet{holoien16d}.

Though some professional surveys have more numerous discoveries, all transients discovered by ASAS-SN are bright and relatively nearby, which allows them to be observed over wide wavelength ranges and long time baselines using fairly modest resources. Spectra of ASAS-SN discoveries are often easily obtainable with a 1-m telescope, and all ASAS-SN supernovae have been spectroscopically confirmed and classified. ASAS-SN also uses an untargeted survey approach, which allows us to provide a less biased tool than other surveys for studying the populations of nearby supernovae and their host galaxies.

This manuscript is the second of a series of yearly catalogs provided by the ASAS-SN team and presents the collected information on supernovae discovered by ASAS-SN in 2015 and their host galaxies. As was done in our 2013 and 2014 catalog \citep{holoien16d}, we provide the same information for bright supernovae, defined as those with $m_{peak}\leq17$, discovered by other professional surveys and amateur astronomers in 2015 in order to construct a full sample of bright supernovae discovered in 2015. The analyses and information presented here supersedes our discovery and classification Astronomer's Telegrams (ATels), all of which are cited in this manuscript, and the information publicly available on the ASAS-SN web pages. We also examine simple statistical properties of the combined supernovae sample from this manuscript and from \citet{holoien16d}. Throughout our analyses, we assume a standard $\Lambda$CDM cosmology with $H_0=69.3$~km~s$^{-1}$~Mpc$^{-1}$, $\Omega_M=0.29$, and $\Omega_{\Lambda}=0.71$ for converting host redshifts into distances.

In \S\ref{sec:sample} we detail the sources of the information presented in this manuscript. In \S\ref{sec:analysis}, we give statistics on the supernovae and hosts in the full, cumulative sample, provide analyses of the data, and discuss the overall trends seen in the sample. Finally, in \S\ref{sec:disc}, we conclude with remarks about the overall findings and look at how future ASAS-SN catalogs will be an integral tool in nearby supernova rate calculations that will be done by the ASAS-SN team.


\section{Data Samples}
\label{sec:sample}

Here we describe the sources of the data collected for our supernova and host galaxy samples, which are presented in Tables~\ref{table:asassn_sne}, \ref{table:other_sne}, \ref{table:asassn_hosts}, and \ref{table:other_hosts}.


\subsection{The ASAS-SN Supernova Sample}
\label{sec:asassn_sample}

The ASAS-SN supernova sample is listed in Table~\ref{table:asassn_sne} and includes all supernovae discovered by ASAS-SN between 2015 January 1 and 2015 December 31. As in \citet{holoien16d}, we collected the names, discovery dates, host names, and host offsets for ASAS-SN discoveries from our discovery Astronomer's Telegrams (ATels), which are cited in Table~\ref{table:asassn_sne}. IAU names are also listed for those supernovae that had one assigned. As we note in our ATels, we view the ASAS-SN names as having priority due to the failure of the IAU system to provide proper credit to the discovery team. We use the IAU system only to avoid temporal reporting conflicts, and encourage others to refer to our discoveries by their ASAS-SN designations.

Redshifts have been spectroscopically measured from classification spectra in all cases. For cases where a redshift had been previously measured for a supernova host galaxy and is consistent with the transient redshift, we list the redshift of the host obtained from the NASA/IPAC Extragalactic Database (NED)\footnote{\url{https://ned.ipac.caltech.edu/}}. For cases where the measurements are not consistent or where a host redshift was not available, we typically report the redshifts given in the classification telegrams. 

Supernova classifications are taken from classification telegrams, which we have also cited in Table~\ref{table:asassn_sne}. For those supernovae that had best-fit ages reported in their classifications, we also give the approximate ages at discovery, measured in days relative to peak. Supernovae were typically classified using either the Supernova Identification code \citep[SNID;][]{blondin07} or the Generic Classification Tool (GELATO\footnote{\url{gelato.tng.iac.es}}; \citealt{harutyunyan08}), both of which compare the observed input spectra to template spectra in order to estimate the best supernova type and age match.

Two supernovae, ASASSN-15da and ASASSN-15hx, never had a redshift announced publicly, and we report redshift measurements here. Using the classification spectrum taken with the Ohio State Multi-Object Spectrograph \citep[OSMOS;][]{martini11} mounted on the MDM Observatory Hiltner 2.4-m telescope we measured a redshift of $0.048\pm0.003$ for ASASSN-15da using SNID.  We measured a redshift of  $0.0081$ for ASASSN-15hx using the host-galaxy H$\alpha$ line detected in a nebular spectrum obtained on 2016 March 11 with the Inamori Magellan Areal Camera and Spectrograph \citep[IMACS;][]{dressler11} mounted on the Magellan-Baade 6.5-m telescope at Las Campanas Observatory.

Based on checking archival classification and late-time spectra of the 2015 ASAS-SN supernova discoveries, we also update a number of redshifts and classifications from the values reported in the discovery and classification telegrams. ASASSN-15al, ASASSN-15az, ASASSN-15bm, ASASSN-15bo, ASASSN-15jm, ASASSN-15lo, ASASSN-15mg, ASASSN-15mx, ASASSN-15nx, ASASSN-15ou, ASASSN-15rg, ASASSN-15sh, ASASSN-15um, ASASSN-15uo, ASASSN-15up, and ASASSN-15ug have updated redshifts that have been remeasured from supernova or host spectra obtained with OSMOS on the MDM 2.4-m telescope, IMACS on the Magellan-Baade 6.5-m telescope, the FAst Spectrograph for the Tillinghast Telescope \citep[FAST;][]{fabricant98}, and the Wide Field Reimaging CCD Camera (WFCCD) mounted on the 2.5-m du Pont telescope at Las Campanas Observatory. Based on a re-examination of the classification spectra, we have updated the classifications of ASASSN-15fa, ASASSN-15fr, ASASSN-15fs, ASASSN-15ga, ASASSN-15hy, ASASSN-15jm, ASASSN-15kg, ASASSN-15kj, ASASSN-15mi, ASASSN-15og, ASASSN-15ou, ASASSN-15tg, ASASSN-15um, ASASSN-15uo, ASASSN-15us, and ASASSN-15ut. We also report an updated redshift for ASASSN-15ed from \citet{pastorello15} and an updated redshift and classification of ASASSN-15no from Benetti et al. (in preparation). These redshifts and classifications are reported in Table~\ref{table:asassn_sne}.

For all ASAS-SN supernovae we solved the astrometry in follow-up images using astrometry.net \citep{barron08} and measured a centroid position for the supernova using IRAF, which typically yielded errors of $<$1\farcs{0} in position. This is significantly more accurate than measuring the coordinates directly from ASAS-SN images, which have a 7\farcs{0} pixel scale. Follow-up images used to measure astrometry were obtained using the LCOGT 1-m telescopes at McDonald Observatory, Cerro Tololo Inter-American Observatory, Siding Springs Observatory, and the South African Astronomical Observatory \citep{brown13}; OSMOS mounted on the MDM Hiltner 2.4-m telescope; the Las Campanas Observatory Swope 1-m telescope; IMACS mounted on the Magellan-Baade 6.5-m telescope; or from amateur collaborators working with the ASAS-SN team. In most cases, we reported coordinates measured from follow-up images in our discovery telegrams, but we report new, more accurate coordinates in Table~\ref{table:asassn_sne} for supernovae that were announced with coordinates measured from ASAS-SN data.

We have re-measured $V$-band, host-subtracted discovery and peak magnitudes from ASAS-SN data for all ASAS-SN supernova discoveries, and report these magnitudes in Table~\ref{table:asassn_sne}. In some cases, re-reductions of the ASAS-SN data have resulted in differences between the magnitudes given in Table~\ref{table:asassn_sne} and those in the original discovery ATels. Here, the ``discovery magnitude'' is defined as the magnitude on the announced discovery date. We also performed a parabolic fit to the measured magnitudes for cases with enough detections, and we use the brighter value between the peak of the fit and the brightest measured magnitude as the ``peak magnitude'' reported in Table~\ref{table:asassn_sne}. For cases with too few detections for a parabolic fit, we use the brightest measured magnitude as the peak magnitude. 

One supernova, ASASSN-15kn, was announced by ASAS-SN in an ATel and was later announced by the Italian Supernovae Search Project\footnote{\url{http://italiansupernovae.org/}} in a the Central Bureau for Astronomical Telegram (CBAT) with the name PSNJ12415045-0710122. The ISSP discovery image was obtained prior to the ASAS-SN discovery image, but since we announced it first, we list it as an ASAS-SN discovery with its ASAS-SN name in Table~\ref{table:asassn_sne}.

All supernovae discovered by ASAS-SN in 2015 are included in this catalog, including those that were fainter than $m_V=17$, but in the comparison analyses presented in \S\ref{sec:analysis} we exclude those supernovae with $m_{V,peak}>17$ so that our sample matches the non-ASAS-SN sample.


\subsection{The Non-ASAS-SN Supernova Sample}
\label{sec:other_sample}

Table~\ref{table:other_sne} gives information for the sample of bright supernovae that were not discovered by ASAS-SN. This sample includes all spectroscopically confirmed supernovae with peak magnitudes $m_{peak}\leq17$ discovered between 2015 January 1 and 2015 December 31. 


\begin{figure*}
\begin{minipage}{\textwidth}
\centering
\subfloat{{\includegraphics[width=0.31\textwidth]{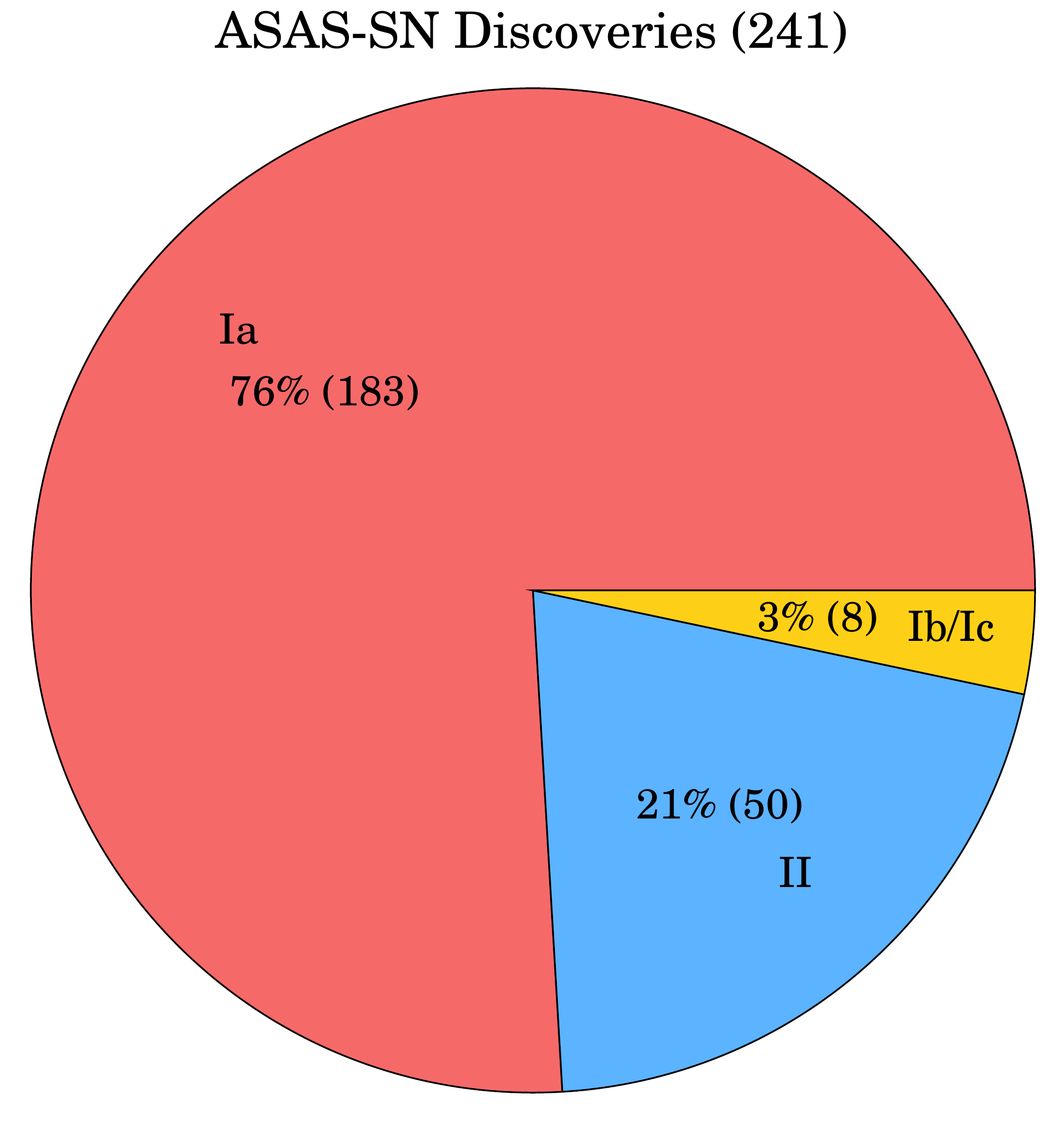}}}
\subfloat{{\includegraphics[width=0.31\textwidth]{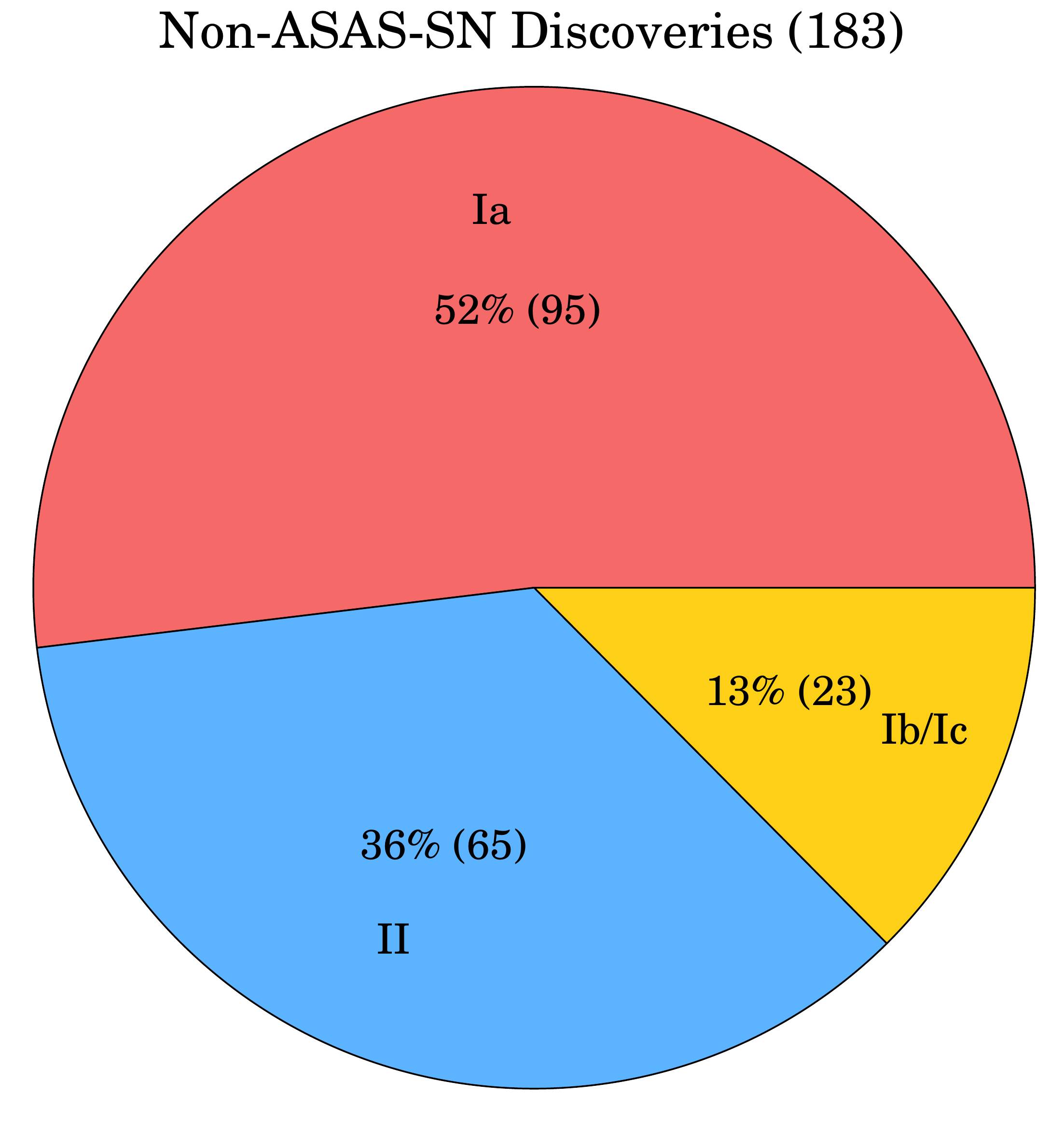}}}
\subfloat{{\includegraphics[width=0.31\textwidth]{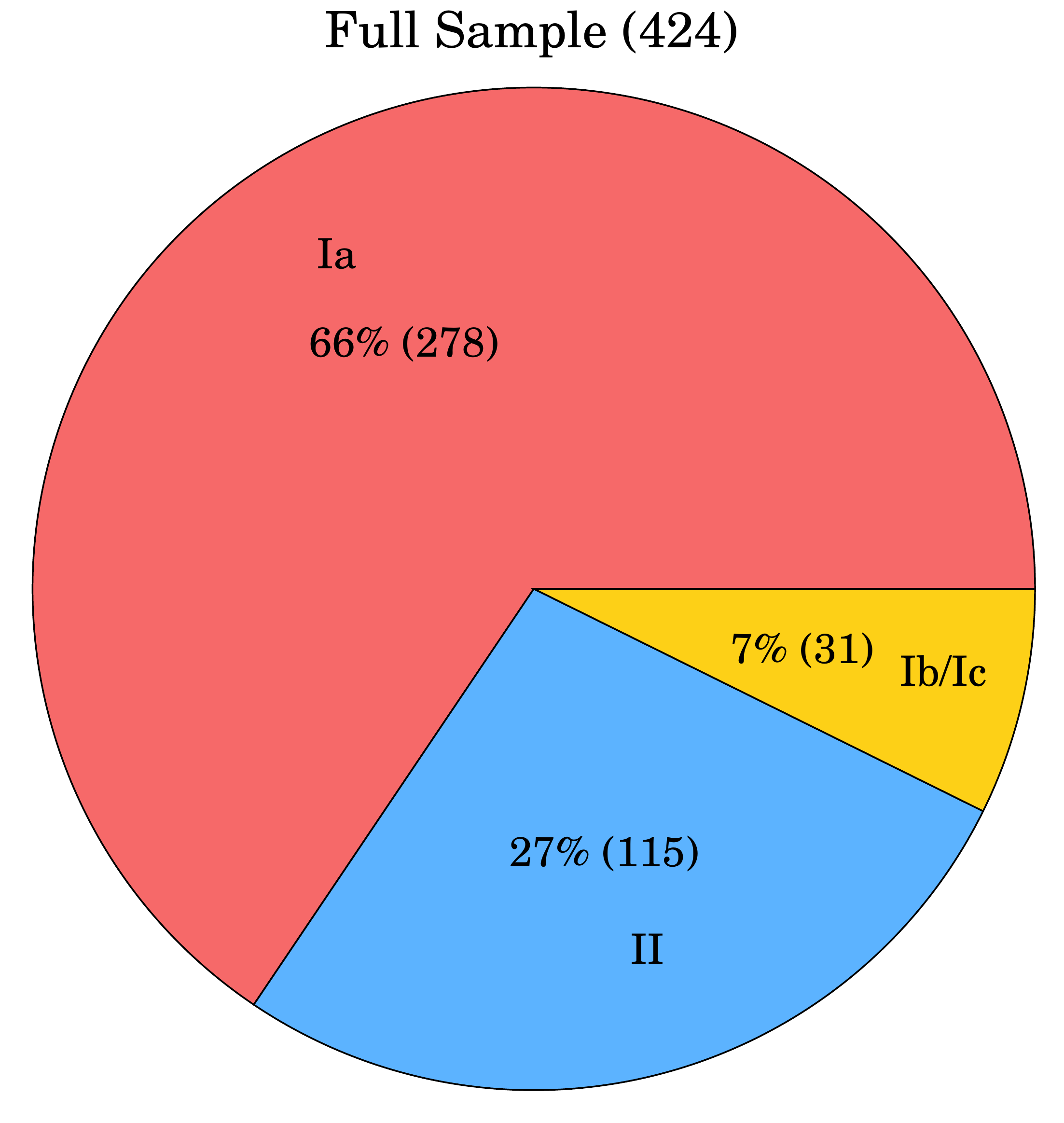}}}
\caption{\emph{Left Panel}: Pie chart breaking down the ASAS-SN sample of supernovae discovered between 2014 May 01 and 2015 December 31 by type. The fractional breakdown is quite similar to that of an ideal magnitude-limited sample from \citet{li11}. \emph{Center Panel}: The same breakdown of supernova types for the Non-ASAS-SN sample from the same time period. \emph{Right Panel:} The same breakdown of supernova types for the combined sample. Type IIb supernovae are considered part of the ``Type II'' sample for this analysis.}
\label{fig:piechart}
\end{minipage}
\end{figure*}

Data for these non-ASAS-SN discoveries were compiled from the ``latest supernovae'' website\footnote{\url{http://www.rochesterastronomy.org/snimages/}} maintained by D. W. Bishop \citep{galyam13}. This page provides a repository of discoveries reported from different channels and attempts to link objects reported by different sources at different times, and thus provides the best available source in 2015 for collecting information on supernovae discovered by different sources. The names, IAU names, discovery dates, coordinates, host names, host offsets, peak magnitudes, types, and discovery sources that are listed in Table~\ref{table:other_sne} were taken from this page when possible. Redshifts for host galaxies were retrieved from NED when available, and were taken from the latest supernovae website when previous host measurements were not available. For cases where the website did not list a host name or host offset for the supernova, this information was taken from NED. We define the offset in these cases as the difference between the reported supernova coordinates and the galaxy coordinates in NED. For all supernovae in both samples, we give the primary name of the host galaxy from NED, which sometimes differs from the name listed on the ASAS-SN supernova page or the latest supernovae website.

We list the name of the discovery group for all supernovae discovered by other professional surveys. For supernovae discovered by non-professional astronomers, we list a discovery source of ``Amateurs'' in order to distinguish these supernovae from those discovered by ASAS-SN and other professional surveys. As was the case in 2014, amateurs accounted for the largest number of bright supernova discoveries after ASAS-SN in 2015.

Lastly, we note in Table~\ref{table:other_sne} whether these supernovae were independently recovered while scanning ASAS-SN data so that we can quantify the impact ASAS-SN would have on the discovery of bright supernovae without the presence of other supernova searches. In \S\ref{sec:missed}, we examine the cases of the two very bright ($m_{peak}<15$) supernovae from 2015 that were not recovered by ASAS-SN in order to determine why ASAS-SN missed these supernovae despite its large coverage area and high cadence.


\subsection{The Host Galaxy Samples}
\label{sec:host_sample}

We have collected Galactic extinction values and magnitudes in various photometric filters spanning from the near-ultraviolet (NUV) to the infrared (IR) for all host galaxies of the supernovae in both the ASAS-SN and the non-ASAS-SN samples. These data are presented in Tables~\ref{table:asassn_hosts} and \ref{table:other_hosts} for ASAS-SN hosts and non-ASAS-SN hosts, respectively. The Galactic $A_V$ is taken from \citet{schlafly11} and was gathered from NED for the supernova position. We obtained NUV magnitudes from the Galaxy Evolution Explorer (GALEX) All Sky Imaging Survey (AIS), optical $ugriz$ magnitudes from the Sloan Digital Sky Survey Data Release 12 \citep[SDSS DR12;][]{alam15}, IR $JHK_S$ magnitudes from the Two-Micron All Sky Survey \citep[2MASS;][]{skrutskie06}, and $W1$ and $W2$ magnitudes from the Wide-field Infrared Survey Explorer \citep[WISE;][]{wright10} AllWISE source catalog. For cases where the host is not detected in 2MASS data, we adopt an upper limit which corresponds to the faintest detected host magnitude in our sample ($m_J>16.5$, $m_H>15.7$) for the $J$- and $H$-bands.

For the $K_S$-band, we estimate a host magnitude for those hosts that are not detected in 2MASS but are detected in the WISE $W1$-band by adding the mean $K_s-W1$ offset from the sample to the WISE $W1$ data. We calculated this offset by averaging the offsets for all hosts that were detected in both the $K_S$- and $W1$-bands from both supernova samples, and it is equal to $-0.64$ magnitudes with a scatter of $0.05$ magnitudes and a standard error of $0.002$ magnitudes. This agrees well with the offset calculated with the 2014 sample in \citet{holoien16d}. For hosts that are not detected in either 2MASS or WISE, we again adopt an upper limit corresponding to the faintest detected host in that filter, $m_{K_S}>15.6$. 


\section{Analysis}
\label{sec:analysis}

Here we provide some basic statistical analyses of the supernova samples and examine reasons why ASAS-SN does not recover some very bright supernovae.


\subsection{Sample Analyses}
\label{sec:sample_anal}


\begin{figure*}
\begin{minipage}{\textwidth}
\centering
\subfloat{{\includegraphics[width=0.7\textwidth]{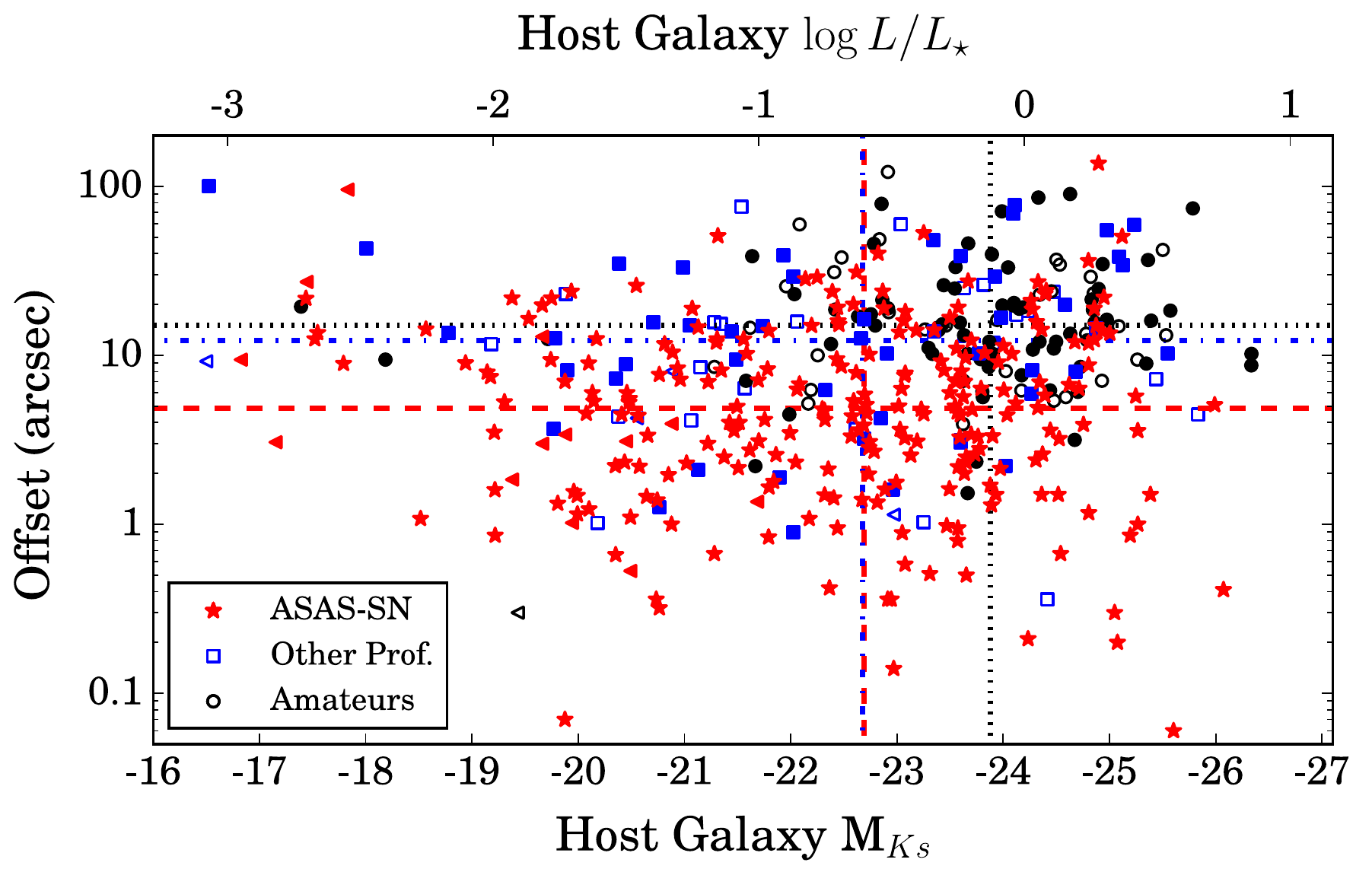}}}

\centering
\subfloat{{\includegraphics[width=0.7\textwidth]{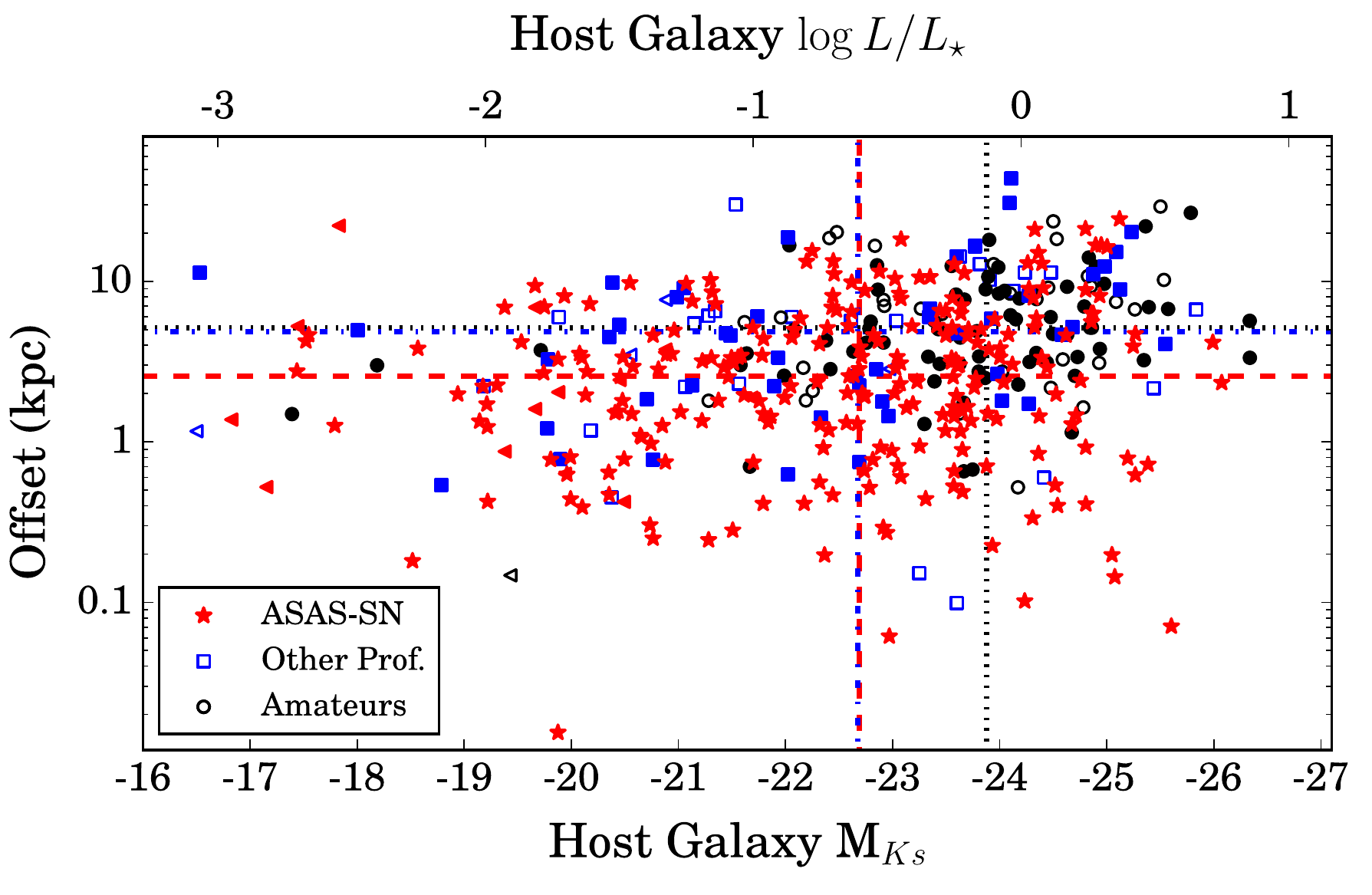}}}
\caption{\emph{Upper Panel}: Absolute $K_S$-band host galaxy magnitude versus the offset from host nucleus in arcseconds for our total bright supernova sample. The upper axis gives the $\log{(L/L_\star)}$ values corresponding to the magnitude scale for $M_{\star,K_S}=-24.2$ \citep{kochanek01}. Supernovae discovered by ASAS-SN are shown as red stars, those discovered by amateur observers are shown as black circles, and those discovered by other professional surveys are shown as blue squares. Filled points indicate supernovae that were independently recovered by ASAS-SN. Triangles are used to indicate upper limits on host galaxy magnitudes for hosts that were not detected by either 2MASS or WISE. Median offsets and magnitudes for ASAS-SN, Amateurs, and other professionals are indicated by dashed, dotted, and dash-dotted lines, respectively, in colors matching the data points. Upper limits are included in computing the median magnitudes. \emph{Lower Panel}: As above, but with the offset measured in kiloparsecs rather than arcseconds.}
\label{fig:offmag}
\end{minipage}
\end{figure*}

Combining all the bright supernovae from both samples discovered between 2014 May 1, when ASAS-SN became operational in both hemispheres, and 2015 December 31 provides a total sample of 425 supernovae, after excluding those ASAS-SN discoveries with $m_{peak}>17.0$. Of these, 57\% (242) were discovered by ASAS-SN, 25\% (105) were discovered by amateurs, and 18\% (78) were discovered by other professional surveys. By type, 278 were Type Ia supernovae, 115 were Type II supernovae, and 31 were Type Ib/Ic supernovae. Following \citet{li11}, we include Type IIb supernovae in the Type II sample to allow a more direct comparison with their results. ASASSN-15lh is either a Type I superluminous supernova \citep[SLSN-I;][]{dong16,godoy-rivera16} or a tidal disruption event around a Kerr black hole \citep{leloudas16}, and we do not include it in analyses that look at trends by type. ASAS-SN discovered 66\% (183) of the Type Ia supernovae, 43\% (50) of the Type II supernovae, and 26\% (8) of the Type Ib/Ic supernovae. Amateurs discovered 20\% (55), 31\% (36), and 48\% (15) of the Type Ia, Type II, and Type Ib/Ic supernovae, respectively, while other professional surveys accounted for the remaining 15\% (41), 25\% (29), and 26\% (8) of each.

In Figure~\ref{fig:piechart} we show a breakdown by type of the supernovae in the ASAS-SN sample, the non-ASAS-SN sample, and the combined sample. As would be expected from a magnitude-limited sample, Type Ia supernovae represent the largest fraction of supernovae in all three samples \citep[e.g.,][]{li11}. The ASAS-SN sample has the highest fraction of Type Ia supernovae of the three samples, and matches the ``ideal magnitude-limited sample'' breakdown (79\% Type Ia, 17\% Type II, and 4\% Type Ib/Ic) predicted from the LOSS volume-limited sample in \citet{li11} almost exactly. As was the case with just the 2014 sample, the non-ASAS-SN sample and the overall sample have higher fractions of Type II and Type Ib/Ic supernovae \citep{holoien16d}.

Although ASAS-SN discovers fewer supernovae overall than other professional surveys, ASAS-SN has been the dominant source of bright supernova discoveries since becoming operational in both hemispheres in May of 2014. Trends seen in the 2014 sample \citep{holoien16d} continue in the 2015 sample as well. ASAS-SN often discovers supernovae shortly after explosion due to its rapid cadence: 217 of the ASAS-SN supernovae have approximate measured ages at discovery, and 70\% (151) were discovered prior to reaching their peak brightness. 

ASAS-SN also continues to be less affected by host galaxy selection effects than other bright supernova searches: 24\% (57) of the ASAS-SN bright supernovae were discovered in catalogued host galaxies without known redshifts, while only 14\% (26) of the supernovae discovered by other sources were found in such hosts. An additional 3\% (7) of the ASAS-SN supernovae were found in uncatalogued hosts or have no apparent host galaxy compared to 1\% (2) of the non-ASAS-SN discoveries, indicating that ASAS-SN is less biased against finding supernovae in uncatalogued hosts.

ASAS-SN discoveries continue to stand out from bright supernovae discovered by other sources with regards to their offsets from their host galaxy nuclei. In Figure~\ref{fig:offmag} we show the host $K_S$-band absolute magnitude and the offset from host nucleus for all supernovae in our sample, with the median values for each source (ASAS-SN, amateurs, or other professionals) marked with horizontal and vertical lines. To help put the overall magnitude scale in perspective, a typical $L_\star$ galaxy has $M_{\star,K_S}=-24.2$ \citep{kochanek01}, and a corresponding luminosity scale is given on the upper axis of the Figure.

Amateur surveys are significantly biased towards luminous galaxies and larger offsets from the host nucleus, which is unsurprising given that they tend to observe bright, nearby galaxies and use less sophisticated detection techniques than professional surveys. This approach allows amateur observers to obtain many observations of such galaxies per night, increasing their chances of finding supernovae, but it biases them against finding supernovae in fainter hosts. Other professional surveys in our comparison sample do discover supernovae with smaller angular separations than amateurs (median value of 12\farcs{2} vs. 15\farcs{0}), but in terms of physical separation both samples exhibit a similar median offset (4.9 kpc for professionals, 5.1 kpc for amateurs). Conversely, ASAS-SN discoveries have median offsets of 4\farcs{9} and 2.6 kpc, indicating that ASAS-SN is less biased against discoveries close to the host nucleus than either comparison group. The three other professional groups with the most discoveries in our comparison sample are CRTS \citep{drake09}, MASTER \citep{gorbovskoy13}, and LOSS \citep{li00}, which either do not use difference imaging or ignore central regions of galaxies in their searches, and this likely contributes to their larger offsets. As we pointed out in \citet{holoien16d}, however, ASAS-SN continues to find a higher rate of tidal disruption events than other surveys (see \citealt{holoien16b,holoien16a}), including those that do use difference imaging, which implies that the avoidance of the central regions of galaxies is still fairly common in surveys other than CRTS, MASTER, and LOSS.


\begin{figure*}
\begin{minipage}{\textwidth}
\centering
\subfloat{{\includegraphics[width=0.75\linewidth]{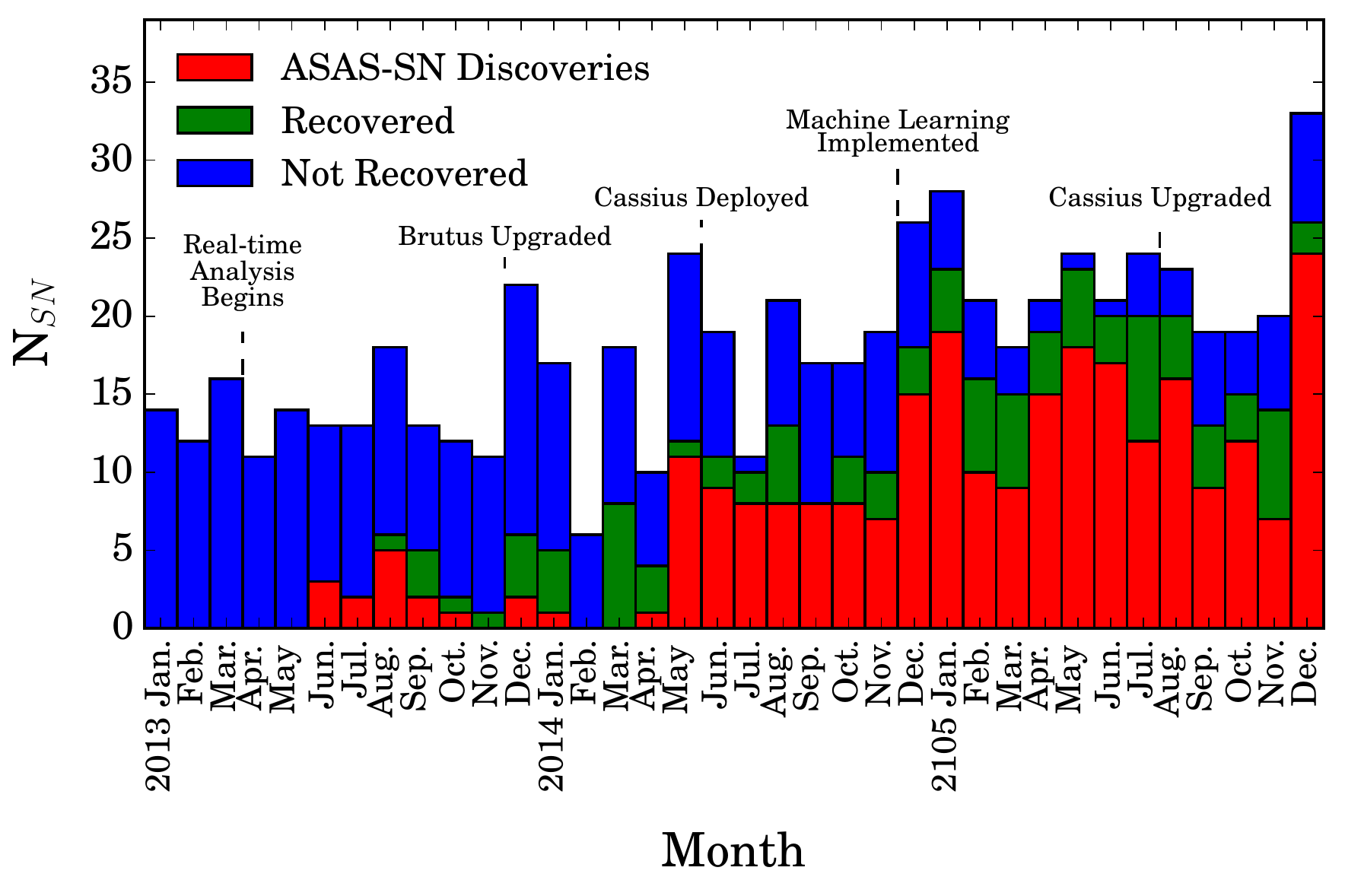}}}
\caption{Histogram of bright supernovae discoveries in each month of 2013, 2014, and 2015. ASAS-SN discoveries are indicated in red, supernovae discovered by other sources and independently recovered by ASAS-SN are indicated in green, and supernovae that were not recovered by ASAS-SN are indicated in blue. Significant milestones in the ASAS-SN timeline are also indicated. The impact of ASAS-SN becoming operational in both hemispheres in 2014 May is clearly seen by a large increase in discoveries in all following months. ASAS-SN accounts for at least half of the bright supernova discoveries in every month after 2014 April, and also recovers a significant fraction of the supernovae it does not discover, particularly in 2015.}
\label{fig:histogram}
\end{minipage}
\end{figure*}

The median host magnitudes for supernovae discovered by ASAS-SN, other professionals, and amateurs are $M_{K_S}\simeq -22.7$, $M_{K_S}\simeq -22.6$, and $M_{K_S}\simeq -23.9$, respectively. While the median host magnitude for other professionals' discoveries is slightly fainter than that for ASAS-SN discoveries, the measurements are consistent given the uncertainties. There is however a clear distinction between professional surveys (including ASAS-SN) and amateurs.

ASAS-SN has had a significant impact on the discovery of bright supernovae. One way this can be seen is by looking at the number of bright supernovae discovered per month in 2013, 2014, and 2015. Figure~\ref{fig:histogram} shows a histogram of supernovae with $m_{peak}\leq17$ discovered by ASAS-SN and those discovered by other sources in each month of 2013, 2014, and 2015. As was the case in \citet{holoien16d}, we include information for all bright supernovae discovered in the early months of 2014 for completeness, and we add the same information for all of 2013 as well to expand our comparison. To help show the impact of various improvements to the ASAS-SN hardware and software, we indicate certain milestones, such as the deployment of our southern unit Cassius and its upgrade to 4 telescopes, on the figure as well.

Before our southern unit Cassius became operational, other supernova searches were discovering the majority of bright, nearby supernovae. However, as can be seen in Figure~\ref{fig:histogram}, the addition of Cassius and improvements to our pipeline had a major impact on our detection efficiency. For every month of 2014 and 2015 after the deployment of Cassius, ASAS-SN discovers or independently recovers at least half of all bright supernovae, and often this fraction is significantly larger. The average number of bright supernovae discovered per month has also increased substantially since ASAS-SN became operational in both hemispheres, from 14 with a scatter of 4 supernovae per month to 21 with a scatter of 5 supernovae per month. This implies that the rate of bright supernovae discovered per month has increased from $\sim13\pm2$ supernovae per month prior to Cassius becoming operational to $\sim21\pm2$ supernovae per month afterwards, providing roughly $4\sigma$ evidence that the discovery rate has increased since Cassius was deployed. This suggests that ASAS-SN is discovering supernovae that would not otherwise be discovered by other supernova searches, and thus that we can now construct a more complete sample of bright, nearby supernovae.

\begin{figure}
\centering
\subfloat{{\includegraphics[width=0.95\linewidth]{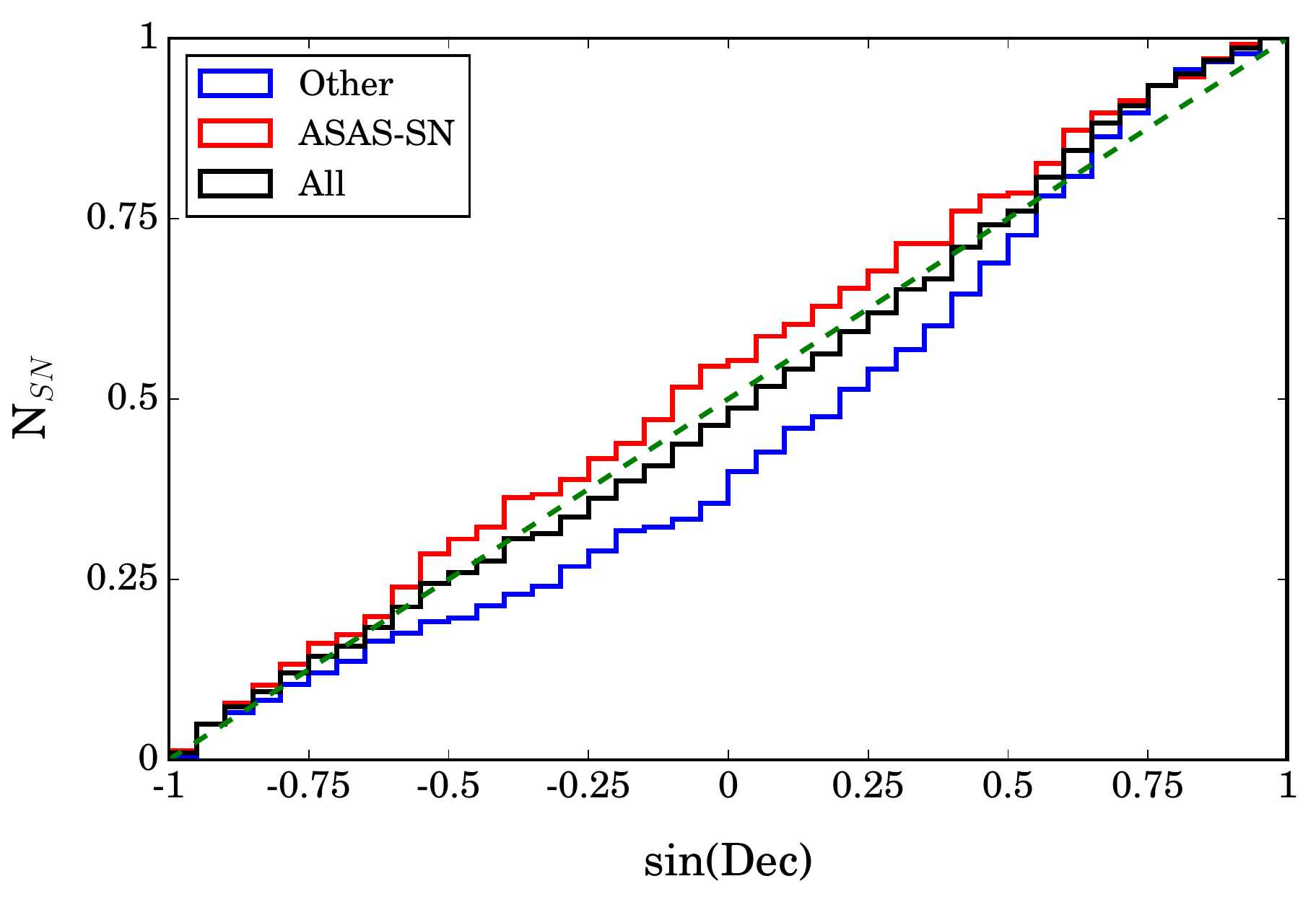}}}
\caption{Cumulative normalized histogram of supernova discoveries with respect to the sine of their declination. ASAS-SN discoveries are shown in red, non-ASAS-SN discoveries are shown in blue, and the combined sample is shown in black. The green dashed line represents what would be expected if supernovae were equally likely to be discovered at all declinations. Non-ASAS-SN discoveries have a clear northern bias and fall well below the expectation except near the poles, while ASAS-SN discoveries correct this trend.}
\label{fig:decplot}
\end{figure}

ASAS-SN is also less biased in terms of the locations of its discoveries. Figure~\ref{fig:decplot} shows a cumulative normalized histogram of supernovae with respect to the sine of their declination. The green dashed line represents what would be expected if supernovae were discovered at all declinations equally. As can be seen in the figure, supernovae discovered by non-ASAS-SN sources have a clear bias towards the northern hemisphere: the non-ASAS-SN histogram falls significantly below the ``no bias'' expectation except at very low ($\sin{\textrm{(Dec)}}\lesssim-0.8$) and very high ($\sin{\textrm{(Dec)}}\gtrsim0.6$) declinations. This is not unexpected, as many professional searches are based in the northern hemisphere. The ASAS-SN discoveries make the combined sample (the black line) follow the expected distribution very closely. The primary bias remaining in sky coverage is the Galactic plane.


\begin{figure}
\centering
\subfloat{{\includegraphics[width=0.95\linewidth]{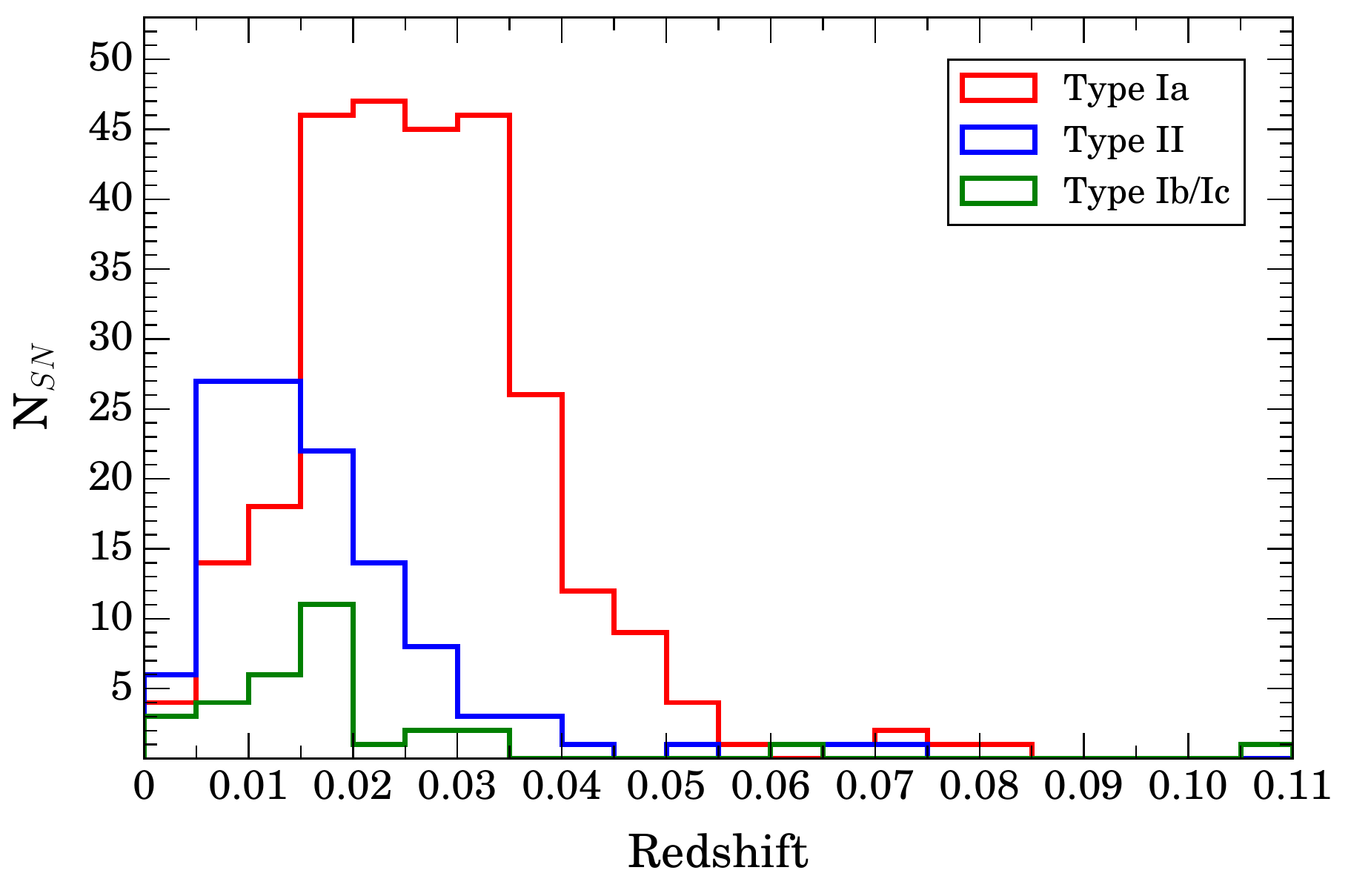}}}
\caption{Histograms of the supernova redshifts with a bin width of $z=0.005$. The red, blue, and green lines show the Type Ia, Type II, and Type Ib/Ic redshift distributions, respectively. Subtypes (e.g., SN 1991T-like Type Ia supernovae) are included as part of their parent groups. Type Ia supernovae in our sample are predominantly found at higher redshifts, while the Type II supernovae are found at comparatively lower redshifts, as expected from a magnitude-limited sample.}
\label{fig:redshift}
\end{figure}

The redshift distribution of our full sample, divided by supernova type, is shown in Figure~\ref{fig:redshift}. There is a clear distinction between the redshifts of Type Ia and Type II supernovae in our sample, with the Type II distribution peaking at $z\lesssim0.01$ and the Type Ia distribution peaking closer to $z\sim0.03$. This distribution is not unexpected for a magnitude-limited sample, as Type Ia supernovae are typically more luminous than core-collapse supernovae. As was the case in our 2014 sample \citep[see][]{holoien16d}, the Type Ib/Ic distribution peaks between $z=0.015$ and $z=0.02$. This trend is not as clear, due in part to the comparatively small number of Type Ib/Ic supernovae in the sample.

\begin{figure}
\centering
\subfloat{{\includegraphics[width=0.95\linewidth]{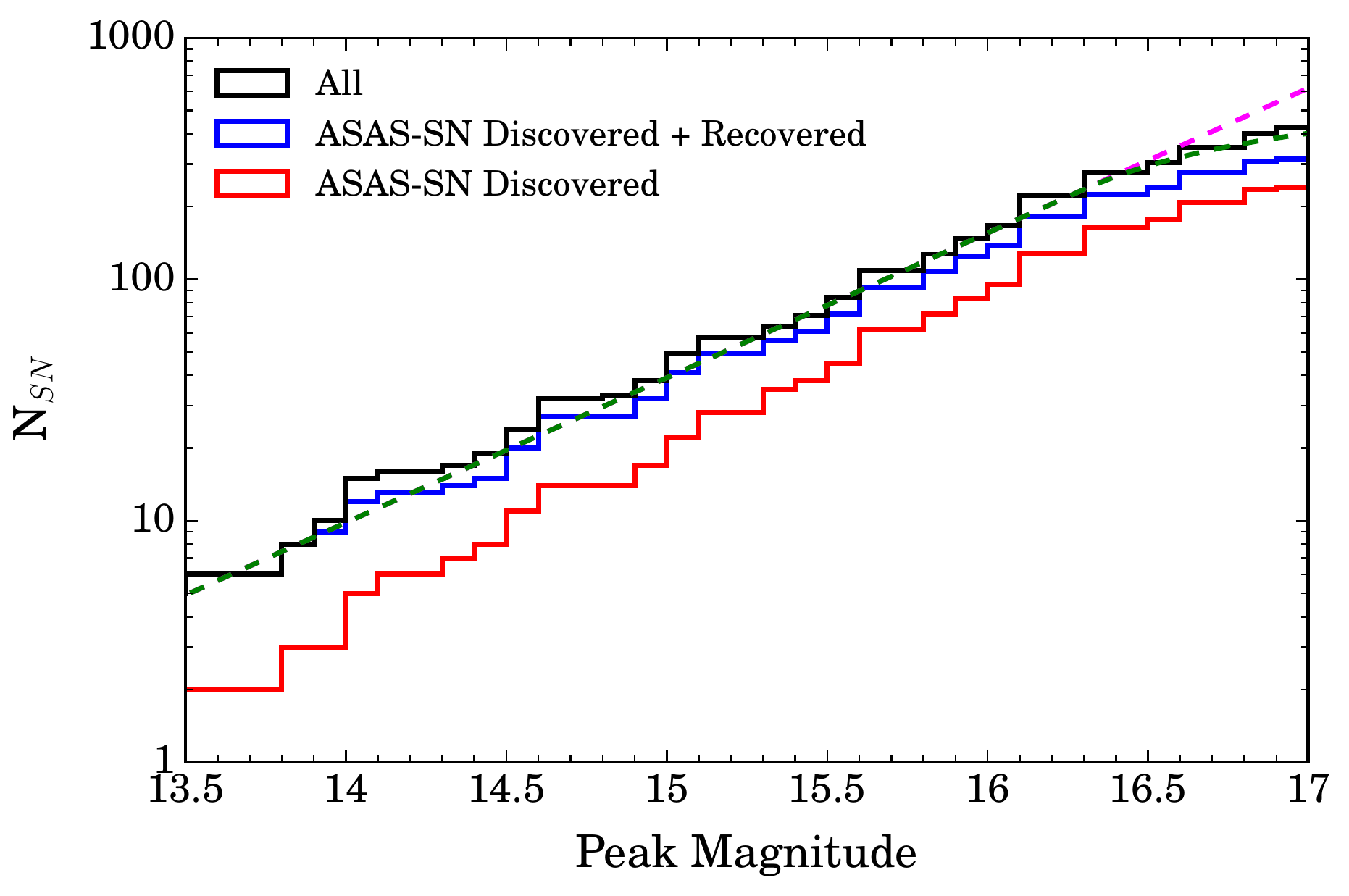}}}
\caption{Cumulative histogram of supernovae with respect to peak magnitude, using a 0.1 magnitude bin width. The black line shows the distribution for all supernovae in our sample, while the red line shows only those supernovae discovered by ASAS-SN, and the blue line includes both ASAS-SN discoveries and those supernovae that were independently recovered in ASAS-SN data. The green dashed line shows a broken power-law fit normalized to the complete sample assuming a Euclidean slope below the break magnitude and a variable slope for fainter sources, while the lavender dashed line indicates an extrapolation of the Euclidean slope to fainter magnitudes. The fits indicate that the sample is roughly $2/3$ complete for $m_{peak}<17$.}
\label{fig:mag_dist}
\end{figure}

Finally, Figure~\ref{fig:mag_dist} shows a cumulative histogram of supernova peak magnitudes with $13.5<m_{peak}<17.0$. ASAS-SN discoveries, ASAS-SN discoveries and recovered supernovae, and all supernovae from our sample are shown in red, blue, and black, respectively. As was the case in the last 8 months of 2014, amateur observers account for a large number of very bright discoveries (those with $m_{peak}\lesssim14.5$; \citealt{holoien16d}). However, in 2015, ASAS-SN discovered a fairly significant fraction of these very bright supernovae as well, showing that it can be competitive with amateurs who observe the relatively small number of very low-redshift galaxies with high cadence. ASAS-SN discovers or recovers every supernova with $m_{peak}<14$, and accounts for a large fraction of the supernovae overall. We discuss the non-recovered cases with $m_{peak}<15$ in \S\ref{sec:missed}. At $m\sim16.5$ the distribution flattens, as such supernovae spend less time at magnitudes bright enough to be found by ASAS-SN. 

A full discussion of supernova rates is being deferred to an upcoming manuscript (Holoien et al., \emph{in prep.}), but Figure~\ref{fig:mag_dist} illustrates the magnitude completeness of our sample. We fit a broken power-law in magnitude (shown with a green dashed line) to the unbinned magnitudes of the theoretically observable SN brighter than $m=17.0$ assuming a Euclidean slope below the break magnitude and a variable slope for fainter sources, deriving the parameters and their uncertainties using Monte Carlo Markov Chain (MCMC) methods.  For the SN discovered by ASAS-SN, the number counts are consistent with a Euclidean slope down to $16.30 \pm 0.08$~mag, which is about 0.3~mag fainter than we found for the 2013/2014 sample and another indication of improvements in our survey and pipelines.  For the samples adding SN later found by ASAS-SN or the sample of all bright SN, we find similar break magnitudes of $16.26\pm0.07$ and $16.27\pm0.07$~mag, respectively.  

The integral completenesses of the three samples relative to the Euclidean predictions are $0.95 \pm 0.03$ ($0.64\pm0.04$), $0.93 \pm 0.03$ ($0.60 \pm 0.03$), and $0.94 \pm 0.03$ ($0.64 \pm0.03$) at 16.5 (17.0)~mag, respectively.  The differential completenesses of the three samples relative to the Euclidean predictions are $0.62 \pm0.09$ ($0.19 \pm 0.05$), $0.54\pm 0.06$ ($0.15 \pm 0.03$) and $0.60 \pm0.07$ ($0.20 \pm 0.04$) at 16.5 (17.0)~mag, respectively.  These results vary little between the three sample definitions and a rough characterization is that $2/3$ of the SN brighter than 17~mag are being found, but only one in $5$-$6$ of the 17~mag SN are being found (relative to the Euclidean extrapolation from the brighter $<16$~mag SN).  The Euclidean approximation modestly underestimates the true completeness for the faint SN  because of deviations from a Euclidean geometry, the effects of time dilation on rates, and K-corrections.  We will include these higher order corrections when we carry out a full analysis of the rates.


\subsection{Examination of Missed Cases}
\label{sec:missed}

In \citet{holoien16d} we performed a retrospective examination of all bright supernovae that were not discovered or recovered by ASAS-SN in a single month (2014 August) in order to better understand some of the reasons ASAS-SN would not recover supernovae that should be bright enough to be detected. Our conclusion was that many of the missed cases in 2014 would be recovered by ASAS-SN today, due to factors such as better cadence, Galactic plane coverage, and an improved detection pipeline.  


This is born out in the observations. From 2014 May 1 through 2014 December 31, ASAS-SN recovered 24\% (19/80) of the supernovae discovered by other sources. In 2015, this fraction increased dramatically, to 55\% (57/104), indicating that our recovery rate has indeed improved since 2014. Of the 47 supernovae that were not recovered, roughly half (23) had peak magnitudes of $m_{peak}\geq16.5$, which makes them more likely to be missed in our data due to factors such as peaking between ASAS-SN observations or occurring during the full moon, when our survey depth is reduced. While it is clear that there is still room for improvement, ASAS-SN is now recovering more than half of the supernovae discovered by other supernova searches, and we expect this fraction to continue to increase in the future.

A similar analysis of missed cases to the one performed in \citet{holoien16d} is unlikely to reveal additional useful information regarding the reasons that ASAS-SN does not recover bright supernovae in our data. We expect that the majority of missed cases in 2015 would have been missed for similar reasons to those discussed in the 2014 catalog. Instead, here we focus on the very bright ($m_{peak}\leq15.0$) supernovae from 2015. These are supernovae that should be detectable in ASAS-SN data regardless of observing conditions, and are also the most interesting supernovae for follow-up study due to their brightness. Understanding the reasons why we missed such supernovae is important to ensure that we miss as few of them as possible going forward. Of the supernovae discovered by other professional searches and amateurs in 2015, 13 had $m_{peak}\leq15.0$. Of these 13, only 2 (SN 2015I and MASTER OT J141023.42-431843.7) were not independently recovered in ASAS-SN data. We examine these two cases in detail here. 

SN 2015I was discovered in NGC 2357 on 2015 May 02 by amateur astronomers (CBET 4106), and peaked at a magnitude of 14.0. Unfortunately, the field containing this supernova was not observable due to Sun constraints from either of our telescope sites for a significant portion of 2015: we were unable to observe the field between 2015 April 22 and 2015 August 29. While the host galaxy had not quite set for the season at the time of discovery, our northern unit Brutus is constrained in how far west it is able to observe due to the presence of the LCOGT 2-m telescope in the same enclosure. Since the supernova was discovered at the beginning of this timeframe, it had faded beyond our detection limit by the time we were able to observe it. While it is unfortunate that we were not able to observe this supernova, there is little that could be done to allow us to discover or recover cases like these.

MASTER OT J141023.42-431843.7 was discovered in NGC 5483 on 2015 December 15 by MASTER \citep{gorbovskoy13}, and peaked at a magnitude of 14.4. This supernova occurred in an observing field that was transferred to one of the new Cassius cameras in the summer of 2015, when we upgraded our Cassius unit to 4 telescopes. Before performing image subtraction in a field, we first construct a reference image by co-adding numerous high-quality exposures. Because of weather and technical issues at the start of operations with four cameras, almost no images were obtained in good conditions prior to the field setting in 2015 September. When the field was again observable in 2016 January, the supernova was present, but we did not have enough high-quality images to build a proper reference image. As a result, new images (containing the supernova) would have been rapidly incorporated into the reference image, preventing detection. The supernova faded below our detection limits on 2016 January 16. Essentially, this was a case of bad timing, as the supernova occurred in a field that was not ready for searching at the time it was observable. The field now has a very good reference image built from images spread over the last year, and the supernova is trivially found in multiple epochs if we analyze the data from 2016 January using the current reference image. 

While we never want to miss any of the brightest supernovae, neither of these cases was observable by ASAS-SN, and the number of very bright supernovae we have missed has decreased over time: in the latter 8 months of 2014 we missed 4 supernovae with $m_{peak}\leq15.0$. This indicates that we have improved our efficiency with very bright supernovae to the point where we are now highly unlikely to miss them in the future if they are observable by ASAS-SN.


\section{Conclusion}
\label{sec:disc}

This manuscript provides the second comprehensive catalog of spectroscopically confirmed bright supernovae and their hosts created by the ASAS-SN team. We now have collected information on two and a half years of ASAS-SN discoveries and 20 months of discoveries from other professional surveys and amateurs. The full combined sample comprises 454 supernovae, 271 of which were discovered by ASAS-SN. We also include an analysis of trends with supernova type, location, peak magnitude, sky position, and redshift, as well as with host galaxy luminosity and supernova offsets from host nuclei. The analyses suggest that the sample closely resembles that of an ideal magnitude-limited survey as described by \citet{li11}, but with a smaller-than-expected proportion of Type Ia supernovae relative to core-collapse supernovae.

ASAS-SN remains the only professional survey attempting to provide a complete, all-sky, rapid-cadence survey of the nearby universe, and as such it has had a significant impact on the discovery and follow-up of bright supernovae in its first 2.5 years of operation. Even today ASAS-SN operates in a region of parameter space that is largely monitored by amateur astronomers, who focus on bright, nearby galaxies for their supernova searches, and it operates with fewer biases. The analysis presented in this paper indicates that ASAS-SN is finding supernovae that otherwise would not be discovered (e.g., Figure~\ref{fig:histogram}), that it eliminates much of the northern hemisphere bias in supernova discoveries (Figure~\ref{fig:decplot}), and that it systematically discovers supernovae closer to galactic nuclei than both amateurs and other professionals and in significantly less luminous galaxies than amateurs (Figure~\ref{fig:offmag}). Compared to our performance in 2014 \citep{holoien16d}, we independently recovered a significantly higher fraction of non-ASAS-SN discoveries in 2015, and we recovered all very bright ($m_{peak}\leq15.0$) supernovae that were observable by our telescopes in 2015. 

The completeness of our full sample has also improved since the end of 2014. In Figure~\ref{fig:mag_dist} we show that the magnitude distribution of the bright supernovae discovered between 2014 May 1 and 2015 December 31 is roughly complete to a peak magnitude of $m_{peak}=16.3$, and is roughly 66\% complete for $m_{peak}\leq17.0$. While we still must address the absolute normalization of the expected number of supernovae (by accounting for factors such as sky coverage and time windows) in order to determine nearby supernovae rates, this analysis serves as a precursor to rate calculations which will be presented in future work by the ASAS-SN team (Holoien et al., \emph{in prep.}). As noted in \citet{holoien16d}, such nearby rate calculations have the potential to impact a number of fields, including the nearby core-collapse rate \citep[e.g.,][]{horiuchi11,horiuchi13} and multi-messenger studies ranging from gravitational waves \citep[e.g.,][]{ando13,nakamura16}, to MeV gamma rays from Type Ia supernovae \citep[e.g.,][]{horiuchi10,diehl14,churazov15} to GeV--TeV gamma rays and neutrinos from rare types of core-collapse supernovae \citep[e.g.,][]{ando05,murase11,abbasi12}. These joint measurements would greatly increase the scientific reach of ASAS-SN discoveries.

This is the second of a yearly series of bright supernova catalogs provided by the ASAS-SN team. It is our hope that by collecting and publishing these data on supernovae and their hosts that we will create convenient and useful repositories that can be used for new and interesting population studies. These catalogs also provide a tool for other professional supernova surveys to perform similar analyses of their data pipelines as we have done for ASAS-SN. While ASAS-SN may discover fewer supernovae than other professional surveys because it is limited to only bright supernovae, it does find the best and the brightest, and these catalogs are one way in which we can fully leverage the unbiased ASAS-SN sample to impact the field of supernova science now and in the future.

\section*{Acknowledgments}

The authors thank LCOGT and its staff for their continued support of ASAS-SN.

ASAS-SN is supported by NSF grant AST-1515927. Development of ASAS-SN has been supported by NSF grant AST-0908816, the Center for Cosmology and AstroParticle Physics at the Ohio State University, the Mt. Cuba Astronomical Foundation, and by George Skestos.

TW-SH is supported by the DOE Computational Science Graduate Fellowship, grant number DE-FG02-97ER25308. JSB, KZS, and CSK are supported by NSF grant AST-1515927. KZS and CSK are also supported by NSF grant AST-1515876. BJS is supported by NASA through Hubble Fellowship grant HST-HF-51348.001 awarded by the Space Telescope Science Institute, which is operated by the Association of Universities for Research in Astronomy, Inc., for NASA, under contract NAS 5-26555. Support for JLP is in part provided by FONDECYT through the grant 1151445 and by the Ministry of Economy, Development, and Tourism's Millennium Science Initiative through grant IC120009, awarded to The Millennium Institute of Astrophysics, MAS. SD is supported by ``the Strategic Priority Research Program-The Emergence of Cosmological Structures'' of the Chinese Academy of Sciences (Grant No. XDB09000000) and Project 11573003 supported by NSFC. JFB is supported by NSF grant PHY-1404311. PRW acknowledges support from the US Department of Energy as part of the Laboratory Directed Research and Development program at LANL.

This research has made use of the XRT Data Analysis Software (XRTDAS) developed under the responsibility of the ASI Science Data Center (ASDC), Italy. At Penn State the NASA {\swift} program is support through contract NAS5-00136.

This research was made possible through the use of the AAVSO Photometric All-Sky Survey (APASS), funded by the Robert Martin Ayers Sciences Fund.

This research has made use of data provided by Astrometry.net \citep{barron08}.

This paper uses data products produced by the OIR Telescope Data Center, supported by the Smithsonian Astrophysical Observatory.

Observations made with the NASA Galaxy Evolution Explorer (GALEX) were used in the analyses presented in this manuscript. Some of the data presented in this paper were obtained from the Mikulski Archive for Space Telescopes (MAST). STScI is operated by the Association of Universities for Research in Astronomy, Inc., under NASA contract NAS5-26555. Support for MAST for non-HST data is provided by the NASA Office of Space Science via grant NNX13AC07G and by other grants and contracts.

Funding for SDSS-III has been provided by the Alfred P. Sloan Foundation, the Participating Institutions, the National Science Foundation, and the U.S. Department of Energy Office of Science. The SDSS-III web site is http://www.sdss3.org/.

This publication makes use of data products from the Two Micron All Sky Survey, which is a joint project of the University of Massachusetts and the Infrared Processing and Analysis Center/California Institute of Technology, funded by NASA and the National Science Foundation.

This publication makes use of data products from the Wide-field Infrared Survey Explorer, which is a joint project of the University of California, Los Angeles, and the Jet Propulsion Laboratory/California Institute of Technology, funded by NASA.

This research has made use of the NASA/IPAC Extragalactic Database (NED), which is operated by the Jet Propulsion Laboratory, California Institute of Technology, under contract with NASA.

\bibliographystyle{mnras2}
\bibliography{bibliography}

\newpage

\begin{landscape}
\begin{table}
\begin{minipage}{\textwidth}
\centering
\fontsize{6}{7.2}\selectfont
\caption{ASAS-SN Supernovae}
\label{table:asassn_sne}
\begin{tabular}{@{}l@{\hspace{0.15cm}}l@{\hspace{0.15cm}}c@{\hspace{0.15cm}}c@{\hspace{0.15cm}}c@{\hspace{0.15cm}}l@{\hspace{0.15cm}}c@{\hspace{0.15cm}}c@{\hspace{0.15cm}}c@{\hspace{0.15cm}}c@{\hspace{0.15cm}}c@{\hspace{0.15cm}}l@{\hspace{0.15cm}}l@{\hspace{0.15cm}}l@{\hspace{-0.05cm}}} 
\hline
\vspace{-0.14cm}
 & & & & & & & & & & & & & \\
 & IAU & Discovery & & & & & & Offset & & Age & & & \\
SN Name & Name$^a$ & Date & RA$^b$ & Dec.$^b$ & Redshift & $V_{disc}$$^c$ & $V_{peak}$$^c$ & (arcsec)$^d$ & Type & at Disc.$^e$ & Host Name & Discovery ATel & Classification ATel \\
\vspace{-0.23cm} \\
\hline
\vspace{-0.17cm}
 & & & & & & & & & & & & &\\
ASASSN-15ab  &  ---  &  2015-01-02.33  &  210:46:35.04  &  $-$38:28:24.6  &  0.01780  &  16.3  &  15.5  &  25.86  &  IIn  &  ---  &  ESO 325-G045  & \citet{asassn15ab_atel} & \citet{asassn15ab_spec_atel} \\ 
ASASSN-15ad  &  ---  &  2014-12-29.18  &  95:56:13.06  &  $-$16:05:51.0  &  0.02400  &  16.3  &  16.3  &  14.67  &  Ia  &  -4  &  PGC 902451  & \citet{asassn15ad_atel} & \citet{asassn15ab_spec_atel} \\ 
ASASSN-15ae  &  ---  &  2015-01-01.63  &  237:46:27.12  &  $+$34:25:55.8  &  0.03124  &  17.4  &  16.4  &  4.98  &  Ia  &  -5  &  2MASX J15510599+3425520  & \citet{asassn15ae_atel} & \citet{asassn15ae_spec_atel} \\ 
ASASSN-15aj  &  ---  &  2015-01-08.30  &  163:13:18.91  &  $-$32:55:34.9  &  0.01092  &  15.0  &  14.7  &  6.36  &  Ia  &  -3  &  NGC 3449  & \citet{asassn15aj_atel} & \citet{asassn15aj_spec_atel} \\ 
ASASSN-15ak  &  ---  &  2015-01-09.22  &  3:00:23.18  &  $+$26:23:37.3  &  0.01503  &  15.1  &  14.7  &  3.36  &  Ia-91T  &  ---  &  UGC 00110  & \citet{asassn15ak_atel} & \citet{asassn15ae_spec_atel} \\ 
ASASSN-15al  &  ---  &  2015-01-10.33  &  74:27:24.44  &  $-$21:35:34.1  &  0.03200  &  17.0  &  16.8  &  7.20  &  Ia  &  0  &  GALEXASC J045749.46  & \citet{asassn15al_atel} & \citet{asassn15al_spec_atel} \\ 
ASASSN-15ar  &  ---  &  2015-01-13.28  &  37:05:05.39  &  $+$10:23:03.8  &  0.02882  &  17.6  &  16.6  &  21.06  &  Ia  &  -6  &  CGCG 439-010  & \citet{asassn15ar_atel} & \citet{asassn15ar_spec_atel} \\ 
ASASSN-15as  &  ---  &  2015-01-13.44  &  144:49:08.26  &  $+$6:25:48.5  &  0.02800  &  16.4  &  16.1  &  3.40  &  Ia  &  -1  &  SDSS J093916.69+062551.1  & \citet{asassn15as_atel} & \citet{asassn15as_spec_atel} \\ 
ASASSN-15az  &  ---  &  2015-01-14.46  &  167:09:23.98  &  $-$10:14:57.8  &  0.02832  &  16.3  &  16.3  &  19.00  &  Ia  &  -1  &  2MASX J11083863-1014456  & \citet{asassn15az_atel} & \citet{asassn15ar_spec_atel} \\ 
ASASSN-15ba  &  ---  &  2015-01-15.64  &  211:13:46.34  &  $+$8:55:14.5  &  0.02313  &  16.6  &  15.8  &  0.86  &  Ia  &  -5  &  SDSS J140455.12+085514.0  & \citet{asassn15ba_atel} & \citet{asassn15ba_spec_atel} \\ 
ASASSN-15bb  &  ---  &  2015-01-16.27  &  195:16:35.69  &  $-$36:36:00.2  &  0.01587  &  16.3  &  16.3  &  10.24  &  II  &  -2  &  ESO 381-IG048  & \citet{asassn15bb_atel} & \citet{asassn15bb_spec_atel} \\ 
ASASSN-15bc  &  ---  &  2015-01-16.31  &  61:33:39.60  &  $-$8:53:08.3  &  0.03672  &  17.1  &  16.9  &  3.64  &  Ia  &  -7  &  2MASX J04061478-0853112  & \citet{asassn15bc_atel} & \citet{asassn15bc_spec_atel} \\ 
ASASSN-15bd  &  ---  &  2015-01-17.63  &  238:39:34.96  &  $+$16:36:38.1  &  0.00795  &  16.8  &  16.0  &  1.08  &  IIb  &  ---  &  SDSS J155438.39+163637.6  & \citet{asassn15bd_atel} & \citet{asassn15bd_spec_atel} \\ 
ASASSN-15be  &  ---  &  2015-01-19.07  &  43:11:35.84  &  $-$34:18:52.5  &  0.02190  &  16.4  &  15.6  &  7.00  &  Ia  &  -10  &  GALEXASC J025245.83  & \citet{asassn15be_atel} & \citet{asassn15be_spec_atel} \\ 
ASASSN-15bf  &  ---  &  2015-01-19.12  &  69:40:16.50  &  $-$41:37:26.2  &  0.04853  &  17.7  &  16.1  &  4.77  &  Ia  &  8  &  2MASX J04384102-4137212  & \citet{asassn15bf_atel} & \citet{asassn15bf_spec_atel} \\ 
ASASSN-15bk  &  ---  &  2015-01-19.62  &  236:03:06.44  &  $+$11:16:16.4  &  0.03428  &  17.2  &  17.1  &  6.99  &  Ia  &  -1  &  CGCG 078-056  & \citet{asassn15bk_atel} & \citet{asassn15be_spec_atel} \\ 
ASASSN-15bm  &  ---  &  2015-01-22.36  &  226:27:53.71  &  $-$5:37:37.1  &  0.02060  &  16.6  &  15.5  &  0.14  &  Ia  &  -10  &  GALEXASC J150551.56  & \citet{asassn15bm_atel} & \citet{asassn15be_spec_atel} \\ 
ASASSN-15bn  &  ---  &  2015-01-23.63  &  238:53:50.24  &  $+$66:43:19.3  &  0.02800  &  16.7  &  16.1  &  7.69  &  Ia  &  -2  &  GALEXASC J155536.57  & \citet{asassn15bn_atel} & \citet{asassn15bn_spec_atel} \\ 
ASASSN-15bo  &  ---  &  2015-01-22.58  &  221:13:48.29  &  $+$24:34:43.7  &  0.03630  &  17.1  &  16.1  &  0.32  &  Ia  &  -9  &  SDSS J144455.21+243443.9  & \citet{asassn15bo_atel} & \citet{asassn15bo_spec_atel} \\ 
ASASSN-15cb  &  ---  &  2015-01-21.54  &  189:57:33.44  &  $+$3:47:49.7  &  0.04004  &  16.8  &  16.8  &  0.51  &  Ia  &  -1  &  VCC 1810  & \citet{asassn15cb_atel} & \citet{asassn15cb_spec_atel} \\
\vspace{-0.22cm}
 & & & & & & & & & & & & &\\
\hline
\end{tabular}
\smallskip
\\
\raggedright
\noindent This table is available in its entirety in a machine-readable form in the online journal. A portion is shown here for guidance regarding its form and content. ``GALEXASC'' galaxy names have been abbreviated for space reasons.\\
$^a$ IAU name is not provided if one was not given to the supernova.\\
$^b$ Right ascension and declination are given in the J2000 epoch. \\
$^c$ All magnitudes are $V$-band magnitudes from ASAS-SN. \\
$^d$ Offset indicates the offset of the SN in arcseconds from the coordinates of the host nucleus, taken from NED. \\
$^e$ Discovery ages are given in days relative to peak. All ages are approximate and are only listed if a clear age was given in the classification telegram. \\
\vspace{-0.5cm}
\end{minipage}
\end{table}


\begin{table}
\begin{minipage}{\textwidth}
\bigskip\bigskip
\centering
\fontsize{6}{7.2}\selectfont
\caption{Non-ASAS-SN Supernovae}
\label{table:other_sne}
\begin{tabular}{@{}l@{\hspace{0.15cm}}l@{\hspace{0.15cm}}c@{\hspace{0.15cm}}c@{\hspace{0.15cm}}c@{\hspace{0.15cm}}l@{\hspace{0.15cm}}c@{\hspace{0.15cm}}c@{\hspace{0.15cm}}c@{\hspace{0.15cm}}l@{\hspace{0.15cm}}c@{\hspace{0.15cm}}c} 
\hline
\vspace{-0.14cm}
 & & & & & & & & & & & \\
 & IAU & Discovery &  & & & & Offset & & & & \\
 SN Name & Name$^{a}$ & Date & RA$^b$ & Dec.$^b$ & Redshift & $m_{peak}$$^c$ & (arcsec)$^d$ & Type & Host Name & Discovered By$^e$ & Recovered?$^f$ \\
\vspace{-0.23cm} \\
\hline
\vspace{-0.17cm}
 & & & & & & & & & & & \\
2015B & 2015B & 2015-01-05.12 & 12:54:35.53 & -12:34:18.6 & 0.01544 & 15.0 & 10.20 & Ia & NGC 4782 & Amateurs & Yes \\
LSQ15z & --- & 2015-01-07.00 & 13:08:22.48 & -8:43:57.8 & 0.07000 & 17.0 & 3.06 & Ia & 2MASX J13082255-0844002 & LSQ & No \\ 
2015C & 2015C & 2015-01-07.60 & 13:18:30.47 & -14:36:44.6 & 0.00964 & 16.4 & 12.60 & II & IC 4221 & LOSS & No \\ 
2015A & 2015A & 2015-01-09.64 & 9:41:15.55 & 35:53:17.4 & 0.02339 & 16.6 & 24.70 & Ia & NGC 2955 & Amateurs & Yes \\ 
PSN J13522411+3941286 & --- & 2015-01-09.90 & 13:52:24.11 & 39:41:28.6 & 0.00722 & 15.6 & 18.65 & IIn & NGC 5337 & Amateurs & Yes \\ 
2015W & 2015W & 2015-01-12.28 & 6:57:43.03 & 13:34:45.7 & 0.01329 & 16.2 & 34.89 & II & UGC 03617 & LOSS & Yes \\ 
2015E & 2015E & 2015-01-13.42 & 3:13:35.31 & 0:15:03.1 & 0.04145 & 16.1 & 1.61 & Ia & KUG 0311+000 & KISS & No \\ 
CSS150120:053013-165338 & --- & 2015-01-20.00 & 5:30:13.08 & -16:53:37.1 & 0.02000 & 17.0 & N/A & II & None & CRTS & No \\ 
PSN J11105565+5322493 & --- & 2015-01-23.14 & 11:10:55.65 & 53:22:49.3 & 0.00955 & 17.0 & 29.15 & II & NGC 3549 & CRTS & No \\ 
PSN J10491665-1938253 & --- & 2015-02-02.00 & 10:49:16.65 & -19:38:25.3 & 0.01403 & 16.8 & 12.04 & IIP & ESO 569-G012 & Amateurs & No \\ 
2015bh & 2015bh & 2015-02-07.39 & 9:09:34.96 & 33:07:20.4 & 0.00649 & 15.4 & 16.40 & IIn/LBV & NGC 2770 & CRTS & Yes \\ 
2015H & 2015H & 2015-02-10.54 & 10:54:42.16 & -21:04:13.8 & 0.01246 & 16.0 & 33.11 & Ia-02cx & NGC 3464 & Amateurs & Yes \\ 
PSN J13471211-2422171 & --- & 2015-02-12.52 & 13:47:12.11 & -24:22:17.1 & 0.01991 & 15.1 & 6.08 & Ia & ESO 509-G108 & Amateurs & Yes \\ 
PSN J10234760+3348477 & --- & 2015-02-13.31 & 10:23:47.60 & 33:48:47.7 & 0.03389 & 17.0 & 23.02 & IIb & UGC 05623 & Amateurs & No \\ 
2015U & 2015U & 2015-02-13.34 & 7:28:53.87 & 33:49:10.6 & 0.01379 & 14.9 & 5.92 & Ibn & NGC 2388 & LOSS & Yes \\ 
iPTF15ku & 2015bq & 2015-02-14.43 & 12:35:06.37 & 31:14:35.4 & 0.02818 & 16.9 & 15.00 & Ia-91T & KUG 1232+315 & PTF & Yes \\ 
CSS150214:140955+173155 & 2015bo & 2015-02-14.44 & 14:09:55.13 & 17:31:55.6 & 0.01620 & 16.8 & 59.08 & Ia-91bg & NGC 5490 & CRTS & Yes \\ 
PSN J19235601-5955321 & --- & 2015-02-16.64 & 19:23:56.01 & -59:55:32.1 & 0.01308 & 15.1 & 18.60 & Ia & NGC 6782 & Amateurs & No \\ 
PSN J20580766-5147074 & --- & 2015-02-22.13 & 20:58:07.66 & -51:47:07.4 & 0.01567 & 16.3 & 10.20 & IIb & FAIRALL 0927 & Amateurs & No \\ 
2015X & 2015X & 2015-02-23.22 & 7:16:42.58 & 29:51:22.7 & 0.01072 & 16.0 & 6.24 & II & UGC 03777 & LOSS & No \\
\vspace{-0.22cm}
 & & & & & & & & & & & \\
\hline
\end{tabular}
\smallskip
\\
\raggedright
\noindent This table is available in its entirety in a machine-readable form in the online journal. A portion is shown here for guidance regarding its form and content.\\
$^a$ IAU name is not provided if one was not given to the supernova. In some cases the IAU name may also be the primary supernova name. \\
$^b$ Right ascension and declination are given in the J2000 epoch. \\
$^c$ All magnitudes are taken from D. W. Bishop's Bright Supernova website, as described in the text, and may be from different filters. \\
$^d$ Offset indicates the offset of the SN in arcseconds from the coordinates of the host nucleus, taken from NED. \\
$^e$ ``Amateurs'' indicates discovery by any number of non-professional astronomers, as described in the text. \\
$^f$ Indicates whether the supernova was independently recovered in ASAS-SN data or not.
\end{minipage}
\vspace{-0.5cm}
\end{table}

\end{landscape}
\pagebreak
\begin{landscape}


\begin{table}
\begin{minipage}{\textwidth}
\centering
\fontsize{6}{7.2}\selectfont
\caption{ASAS-SN Supernova Host Galaxies}
\label{table:asassn_hosts}
\begin{tabular}{@{}l@{\hspace{0.15cm}}l@{\hspace{0.15cm}}c@{\hspace{0.15cm}}c@{\hspace{0.15cm}}c@{\hspace{0.15cm}}c@{\hspace{0.15cm}}c@{\hspace{0.15cm}}c@{\hspace{0.15cm}}c@{\hspace{0.15cm}}c@{\hspace{0.15cm}}c@{\hspace{0.15cm}}c@{\hspace{0.15cm}}c@{\hspace{0.15cm}}c@{\hspace{0.15cm}}c@{\hspace{0.15cm}}c@{\hspace{0.15cm}}c} 
\hline
\vspace{-0.14cm}
 & & & & & & & & & & & & & & \\
 & & SN & SN & SN Offset & & & & & & & & & & \\
Galaxy Name & Redshift & Name & Type & (arcsec) & $A_V$$^a$ & $m_{NUV}$$^b$ & $m_u$$^c$ & $m_g$$^c$ & $m_r$$^c$ & $m_i$$^c$ & $m_z$$^c$ & $m_J$$^d$ & $m_H$$^d$ & $m_{K_S}$$^{d,e}$ & $m_{W1}$ & $m_{W2}$\\ 
\vspace{-0.23cm} \\
\hline
\vspace{-0.17cm}
 & & & & & & & & & & & & & & \\
ESO 325-G045 & 0.01780 & ASASSN-15ab & IIn & 27.03 & 0.301 & 16.36 0.02 & --- & --- & --- & --- & --- & 14.81 0.08 & 14.10 0.09 & 13.95 0.16 & 13.08 0.03 & 12.97 0.03 \\ 
PGC 902451 & 0.02400 & ASASSN-15ad & Ia & 14.67 & 0.577 & --- & --- & --- & --- & --- & --- & $>$16.5 & $>$15.7 & 14.06 0.06* & 14.60 0.03 & 14.64 0.05 \\ 
2MASX J15510599+3425520 & 0.03124 & ASASSN-15ae & Ia & 4.98 & 0.085 & 17.62 0.03 & 17.56 0.02 & 16.27 0.00 & 15.83 0.00 & 15.59 0.00 & 15.45 0.01 & 15.02 0.09 & 14.66 0.16 & 14.22 0.15 & 14.26 0.03 & 14.11 0.04 \\ 
NGC 3449 & 0.01092 & ASASSN-15aj & Ia & 6.36 & 0.209 & 16.17 0.03 & --- & --- & --- & --- & --- & 9.70 0.02 & 9.00 0.02 & 8.70 0.03 & 9.60 0.02 & 9.57 0.02 \\ 
UGC 00110 & 0.01503 & ASASSN-15ak & Ia-91T & 3.6 & 0.118 & 17.00 0.01 & 16.66 0.02 & 15.36 0.00 & 14.88 0.00 & 14.67 0.00 & 14.61 0.01 & $>$16.5 & $>$15.7 & 13.45 0.06* & 13.99 0.03 & 13.85 0.04 \\ 
GALEXASC J045749.46 & 0.03200 & ASASSN-15al & Ia & 7.2 & 0.094 & 18.91 0.07 & --- & --- & --- & --- & --- & $>$16.5 & $>$15.7 & 14.81 0.06* & 15.35 0.04 & 15.08 0.06 \\ 
CGCG 439-010 & 0.02882 & ASASSN-15ar & Ia & 21.06 & 0.299 & --- & --- & --- & --- & --- & --- & 12.33 0.03 & 11.66 0.04 & 11.30 0.07 & 12.15 0.02 & 12.11 0.02 \\ 
SDSS J093916.69+062551.1 & 0.02800 & ASASSN-15as & Ia & 3.4 & 0.124 & 21.73 0.14 & 21.02 0.23 & 19.85 0.04 & 19.61 0.05 & 19.44 0.06 & 19.56 0.22 & $>$16.5 & $>$15.7 & $>$15.6 & --- & --- \\ 
2MASX J11083863-1014456 & 0.02832 & ASASSN-15az & Ia & 19 & 0.171 & 21.80 0.46 & --- & --- & --- & --- & --- & 13.76 0.06 & 13.03 0.05 & 12.63 0.10 & 12.98 0.04 & 13.02 0.04 \\ 
SDSS J140455.12+085514.0 & 0.02313 & ASASSN-15ba & Ia & 0.86 & 0.070 & 19.76 0.04 & 19.19 0.10 & 18.04 0.02 & 17.59 0.02 & 17.36 0.02 & 17.21 0.06 & $>$16.5 & $>$15.7 & 15.83 0.08* & 16.37 0.06 & 16.14 0.17 \\ 
ESO 381-IG048 & 0.01587 & ASASSN-15bb & II & 10.24 & 0.130 & 16.20 0.02 & --- & --- & --- & --- & --- & $>$16.5 & $>$15.7 & 12.64 0.05* & 13.18 0.02 & 12.95 0.03 \\ 
2MASX J04061478-0853112 & 0.03672 & ASASSN-15bc & Ia & 3.64 & 0.119 & --- & --- & --- & --- & --- & --- & 14.02 0.06 & 13.36 0.07 & 13.05 0.11 & 13.11 0.03 & 12.95 0.03 \\ 
SDSS J155438.39+163637.6 & 0.00795 & ASASSN-15bd & IIb & 1.08 & 0.094 & --- & 17.70 0.02 & 16.99 0.00 & 16.65 0.00 & 16.57 0.01 & 16.48 0.01 & $>$16.5 & $>$15.7 & 14.19 0.06* & 14.73 0.03 & 14.60 0.05 \\ 
GALEXASC J025245.83 & 0.02190 & ASASSN-15be & Ia & 7 & 0.054 & 18.61 0.05 & --- & --- & --- & --- & --- & $>$16.5 & $>$15.7 & 15.05 0.06* & 15.59 0.04 & 15.41 0.07 \\ 
2MASX J04384102-4137212 & 0.04853 & ASASSN-15bf & Ia & 4.77 & 0.069 & --- & --- & --- & --- & --- & --- & 13.77 0.06 & 13.21 0.08 & 12.99 0.13 & 12.62 0.09 & 12.53 0.02 \\ 
CGCG 078-056 & 0.03428 & ASASSN-15bk & Ia & 6.99 & 0.125 & 18.23 0.02 & 17.30 0.02 & 15.80 0.00 & 15.02 0.00 & 14.61 0.00 & 14.27 0.01 & 13.45 0.05 & 12.73 0.05 & 12.37 0.09 & 12.54 0.02 & 12.38 0.02 \\ 
GALEXASC J150551.56 & 0.02060 & ASASSN-15bm & Ia & 0.14 & 0.221 & 17.87 0.05 & --- & --- & --- & --- & --- & 12.92 0.05 & 12.35 0.07 & 11.84 0.09 & 12.21 0.02 & 12.20 0.02 \\ 
GALEXASC J155536.57 & 0.02800 & ASASSN-15bn & Ia & 7.69 & 0.086 & 18.83 0.06 & --- & --- & --- & --- & --- & $>$16.5 & $>$15.7 & 14.71 0.06* & 15.25 0.03 & 15.19 0.05 \\ 
SDSS J144455.21+243443.9 & 0.03630 & ASASSN-15bo & Ia & 0.32 & 0.116 & 20.67 0.28 & 19.73 0.05 & 18.70 0.01 & 18.28 0.01 & 18.05 0.01 & 17.86 0.03 & $>$16.5 & $>$15.7 & 15.29 0.06* & 15.83 0.04 & 16.00 0.12 \\ 
VCC 1810 & 0.04004 & ASASSN-15cb & Ia & 0.51 & 0.090 & 17.56 0.01 & 16.79 0.01 & 15.64 0.00 & 15.14 0.00 & 14.86 0.00 & 14.67 0.01 & 13.73 0.06 & 12.96 0.06 & 12.96 0.12 & 12.75 0.02 & 12.56 0.03 \\
\vspace{-0.22cm}
 & & & & & & & & & & & & & & \\
\hline
\end{tabular}
\smallskip
\\
\raggedright
\noindent This table is available in its entirety in a machine-readable form in the online journal. A portion is shown here for guidance regarding its form and content. Uncertainty is given for all magnitudes, and in some cases is equal to zero. ``GALEXASC'' galaxy names have been abbreviated for space reasons.\\
$^a$ Galactic extinction taken from \citet{schlafly11}. \\
$^b$ No magnitude is listed for those galaxies not detected in GALEX survey data. \\
$^c$ No magnitude is listed for those galaxies not detected in SDSS data or those located outside of the SDSS footprint. \\
$^d$ For those galaxies not detected in 2MASS data, we assume an upper limit of the faintest galaxy detected in each band from our sample. \\
$^e$ $K_S$-band magnitudes marked with a ``*'' indicate those estimated from the WISE $W1$-band data, as described in the text. \\
\end{minipage}
\vspace{-0.5cm}
\end{table}


\begin{table}
\begin{minipage}{\textwidth}
\bigskip\bigskip
\centering
\fontsize{6}{7.2}\selectfont
\caption{Non-ASAS-SN Supernova Host Galaxies}
\label{table:other_hosts}
\begin{tabular}{@{}l@{\hspace{0.15cm}}l@{\hspace{0.15cm}}c@{\hspace{0.15cm}}c@{\hspace{0.05cm}}c@{\hspace{0.15cm}}c@{\hspace{0.15cm}}c@{\hspace{0.15cm}}c@{\hspace{0.15cm}}c@{\hspace{0.15cm}}c@{\hspace{0.15cm}}c@{\hspace{0.15cm}}c@{\hspace{0.15cm}}c@{\hspace{0.15cm}}c@{\hspace{0.15cm}}c@{\hspace{0.15cm}}c@{\hspace{0.15cm}}c} 
\hline
\vspace{-0.14cm}
 & & & & & & & & & & & & & & \\
 & & SN & SN & SN Offset & & & & & & & & & & \\
Galaxy Name & Redshift & Name & Type & (arcsec) & $A_V$$^a$ & $m_{NUV}$$^b$ & $m_u$$^c$ & $m_g$$^c$ & $m_r$$^c$ & $m_i$$^c$ & $m_z$$^c$ & $m_J$$^d$ & $m_H$$^d$ & $m_{K_S}$$^{d,e}$ & $m_{W1}$ & $m_{W2}$ \\ 
\vspace{-0.23cm} \\
\hline
\vspace{-0.17cm}
 & & & & & & & & & & & & & & \\
NGC 4782 & 0.01544 & 2015B & Ia & 10.20 & 0.146 & 16.65 0.01 & --- & --- & --- & --- & --- & 8.82 0.01 & 8.10 0.01 & 7.83 0.02 & 9.48 0.02 & 9.54 0.02 \\ 
2MASX J13082255-0844002 & 0.07000 & LSQ15z & Ia & 3.06 & 0.098 & 19.57 0.07 & --- & --- & --- & --- & --- & 15.05 0.09 & 14.32 0.10 & 13.93 0.15 & 13.54 0.03 & 13.16 0.03 \\ 
IC 4221 & 0.00964 & 2015C & II & 12.60 & 0.223 & --- & --- & --- & --- & --- & --- & 11.41 0.03 & 10.72 0.03 & 10.48 0.06 & 11.65 0.02 & 11.61 0.02 \\ 
NGC 2955 & 0.02339 & 2015A & Ia & 24.70 & 0.030 & 15.71 0.02 & 14.69 0.00 & 13.22 0.00 & 12.60 0.00 & 12.19 0.00 & 11.89 0.00 & 11.16 0.02 & 10.47 0.04 & 10.17 0.03 & 10.67 0.02 & 10.52 0.02 \\ 
NGC 5337 & 0.00722 & PSN J13522411 & IIn & 18.65 & 0.039 & 16.47 0.01 & 14.82 0.01 & 13.21 0.00 & 12.49 0.00 & 12.12 0.00 & 11.81 0.00 & 10.97 0.02 & 10.32 0.02 & 10.07 0.03 & 11.13 0.02 & 11.19 0.02 \\ 
UGC 03617 & 0.01329 & 2015W & II & 34.89 & 0.380 & --- & --- & --- & --- & --- & --- & 14.18 0.05 & 13.56 0.07 & 13.48 0.10 & 13.69 0.03 & 13.58 0.03 \\ 
KUG 0311+000 & 0.04145 & 2015E & Ia & 1.61 & 0.229 & 19.28 0.02 & 18.21 0.02 & 16.82 0.00 & 16.15 0.00 & 15.77 0.00 & 15.47 0.01 & 14.29 0.04 & 13.59 0.05 & 13.40 0.07 & 13.07 0.03 & 12.93 0.03 \\ 
None & 0.02000 & CSS150120:053013 & II & N/A & 0.193 & --- & --- & --- & --- & --- & --- & $>$16.5 & $>$15.7 & 15.94 0.15* & 16.58 0.14 & 16.70 0.35 \\ 
NGC 3549 & 0.00955 & PSN J11105565 & II & 29.15 & 0.035 & 15.19 0.00 & 14.61 0.01 & 12.82 0.00 & 12.04 0.00 & 11.67 0.00 & 11.41 0.00 & 10.15 0.02 & 9.45 0.02 & 9.18 0.03 & 11.45 0.03 & 11.51 0.02 \\ 
ESO 569-G012 & 0.01403 & PSN J10491665 & IIP & 12.04 & 0.111 & 16.79 0.02 & --- & --- & --- & --- & --- & 10.67 0.02 & 9.93 0.02 & 9.61 0.03 & 9.85 0.02 & 9.69 0.02 \\ 
NGC 2770 & 0.00649 & 2015bh & IIn/LBV & 16.40 & 0.062 & 14.79 0.01 & --- & --- & --- & --- & --- & 10.52 0.02 & 9.81 0.03 & 9.57 0.04 & 11.64 0.02 & 11.59 0.02 \\ 
NGC 3464 & 0.01246 & 2015H & Ia-02cx & 33.11 & 0.150 & 14.99 0.01 & 15.22 0.01 & 13.45 0.00 & 12.61 0.00 & 12.18 0.00 & 11.87 0.00 & 10.61 0.03 & 9.89 0.03 & 9.65 0.05 & 11.27 0.02 & 11.27 0.02 \\ 
ESO 509-G108 & 0.01991 & PSN J13471211 & Ia & 6.08 & 0.203 & 19.17 0.10 & --- & --- & --- & --- & --- & 10.99 0.02 & 10.29 0.03 & 10.03 0.03 & 10.44 0.02 & 10.53 0.02 \\ 
UGC 05623 & 0.03389 & PSN J10234760 & IIb & 23.02 & 0.067 & --- & 16.88 0.02 & 15.78 0.00 & 15.37 0.00 & 15.18 0.00 & 15.02 0.01 & $>$16.5 & $>$15.7 & 13.76 0.07* & 14.40 0.03 & 14.09 0.04 \\ 
NGC 2388 & 0.01379 & 2015U & Ibn & 5.92 & 0.159 & 18.00 0.03 & 16.21 0.01 & 14.28 0.00 & 13.12 0.00 & 12.56 0.00 & 12.05 0.00 & 10.83 0.01 & 10.03 0.02 & 9.65 0.02 & 9.63 0.02 & 9.16 0.02 \\ 
KUG 1232+315 & 0.02818 & iPTF15ku & Ia-91T & 15.00 & 0.037 & 18.45 0.00 & 18.11 0.03 & 17.01 0.01 & 16.55 0.01 & 16.31 0.01 & 16.12 0.02 & $>$16.5 & $>$15.7 & 14.33 0.07* & 14.97 0.03 & 14.72 0.06 \\ 
NGC 5490 & 0.01620 & CSS150214:140955 & Ia-91bg & 59.08 & 0.073 & 17.92 0.04 & 14.77 0.01 & 12.75 0.00 & 11.91 0.00 & 11.49 0.00 & 11.21 0.00 & 10.01 0.01 & 9.30 0.02 & 9.03 0.02 & 9.71 0.02 & 9.75 0.02 \\ 
NGC 6782 & 0.01308 & PSN J19235601 & Ia & 18.60 & 0.163 & 15.48 0.00 & --- & --- & --- & --- & --- & 9.90 0.01 & 9.21 0.02 & 8.96 0.02 & 9.62 0.02 & 9.54 0.02 \\ 
FAIRALL 0927 & 0.01567 & PSN J20580766- & IIb & 10.20 & 0.099 & --- & --- & --- & --- & --- & --- & 11.86 0.03 & 11.12 0.04 & 10.86 0.05 & 11.08 0.02 & 10.85 0.02 \\ 
UGC 03777 & 0.01072 & 2015X & II & 6.24 & 0.162 & --- & 21.74 0.17 & 20.75 0.07 & 19.36 0.02 & 19.05 0.02 & 18.76 0.05 & 12.03 0.04 & 11.33 0.05 & 11.04 0.07 & 11.93 0.02 & 11.86 0.02 \\
\vspace{-0.22cm}
 & & & & & & & & & & & & & & \\
\hline
\end{tabular}
\smallskip
\\
\raggedright
\noindent This table is available in its entirety in a machine-readable form in the online journal. A portion is shown here for guidance regarding its form and content. Uncertainty is given for all magnitudes, and in some cases is equal to zero. ``PSN'' and ``CSS'' supernova names have been abbreviated for space reasons.\\
$^a$ Galactic extinction taken from \citet{schlafly11}. \\
$^b$ No magnitude is listed for those galaxies not detected in GALEX survey data. \\
$^c$ No magnitude is listed for those galaxies not detected in SDSS data or those located outside of the SDSS footprint. \\
$^d$ For those galaxies not detected in 2MASS data, we assume an upper limit of the faintest galaxy detected in each band from our sample. \\
$^e$ $K_S$-band magnitudes marked with a ``*'' indicate those estimated from the WISE $W1$-band data, as described in the text. \\
\end{minipage}
\vspace{-0.5cm}
\end{table}

\end{landscape}

\end{document}